\pdfoutput=1
\documentclass[%
 reprint,
 amsmath,amssymb,
 aps,
]{revtex4-1}

\usepackage{graphicx}
\usepackage{dcolumn}
\usepackage{bm}

\usepackage[caption=false]{subfig}
\usepackage{multirow}

\begin{document}

\title{Dynamical bunching and density peaks in expanding Coulomb clouds}
\author{B. S. Zerbe}\email{zerbe@msu.edu}
\author{X. Xiang}
\author{C.-Y. Ruan}
\author{S. M. Lund}
\author{P. M. Duxbury}\email{duxbury@msu.edu}

\affiliation{Department of Physics and Astronomy, Michigan State University}
\date{\today}

\newcommand{\vect}[1]{\boldsymbol{#1}}​

\begin{abstract}
Expansion dynamics of single-species, non-neutral clouds, such as electron bunches used in ultrafast electron microscopy, show novel behavior due to high acceleration of
particles in the cloud interior. This often leads to electron bunching and dynamical formation of a density shock in the outer regions of the bunch. 
We develop analytic fluid models to capture these effects, and the analytic predictions are validated by PIC and N-particle simulations. 
In the space-charge dominated regime, two and three dimensional systems with Gaussian initial densities show bunching and a strong shock response, 
while one dimensional systems do not; moreover these effects can be tuned using the initial particle density profile and velocity chirp.
\end{abstract}
\maketitle


\section{Introduction}
Non-neutral plasma systems arise in a variety of physical contexts ranging from 
astrophysics\cite{Arbanil:2014_charged_star_review,Maurya:2015_charged_sphere,Yousefi:2014_dust_aggregates}; 
accelerator technologies \cite{Bacci:2014_plasma_acceleration,Boine:2015_intense_beams,Whelan:2016_MRI,Bernal:2016_recirculator};
ion and neutron production \cite{Bulanov:2002_charged_beam_generation,Fukuda:2009_species_generation,Esirkepov:2004_highly_efficient_ion_generation,Kaplan:2015_preprint,Parks:2001_neutron_production,Bychenkov_2015_review}; 
sources for electron and ion microscopy\cite{Murphy:2014_cold_ions,Gahlmann:2008_ultrashort};
to high power vacuum electronics\cite{Booske:2011_vacuum_review,Liu:2015_maximal_charge,Zhang:2016_review}.  Understanding of the dynamics of spreading
of such systems is critical to the design of next generation technologies, and simple analytic models are particularly helpful for instrument design. 
As a result, substantial theoretical efforts have already been made in this 
vein\cite{Jansen:1988_book,Reiser:1994_book,Batygin:2001_self,Bychenkov:2005_coulomb_explosion,Grech:2011_coulomb_explosion,Kaplan:2003_shock,Kovalev:2005_kinetic_spherically_coulomb_explosion,Last:1997_analytic_coulomb_explosion,Eloy:2001_coulomb_explosion,Krainov:2001_ce_dynamics,Morrison:2015_slow_down_dynamics,Boella:2016_multiple_species}. 
Specifically, free expansion of clouds of charged single-specie 
particles starting from rest have been well studied both analytically
and computationally\cite{Last:1997_analytic_coulomb_explosion,Eloy:2001_coulomb_explosion,Grech:2011_coulomb_explosion,Batygin:2001_self,Degtyareva:1998_gaussian_pileup,Siwick:2002_mean_field,Qian:2002_fluid_flow,Reed:2006_short_pulse_theory,Collin:2005_broadening,Gahlmann:2008_ultrashort,Tao:2012_space_charge,Portman:2013_computational_characterization,Portman:2014_image_charge,Michalik:2006_analytic_gaussian}, 
and a number of studies have found evidence of the formation of a region of high-density, often termed a 
``shock'', on the periphery of the clouds under certain 
conditions\cite{Grech:2011_coulomb_explosion,Kaplan:2003_shock,Kovalev:2005_kinetic_spherically_coulomb_explosion,Last:1997_analytic_coulomb_explosion,Murphy:2014_cold_ions,Reed:2006_short_pulse_theory,Degtyareva:1998_gaussian_pileup}.

One application of these theories that is of particular current interest is to high-density electron clouds used in next-generation ultrafast electron microscopy (UEM) 
development\cite{King:2005_review,Hall:2014_report,Williams:2017_longitudinal_emittance}. 
The researchers in the UEM and the ultrafast electron diffraction (UED) communities have 
conducted substantial theoretical treatment of initially extremely short bunches of 
thousands to ultimately hundreds of millions of 
electrons that operate in a regime dominated by a virtual cathode (VC) 
limit\cite{Valfells:2002_vc_limit,Luiten:2004_uniform_ellipsoidal,King:2005_review,Miller:2014_science_review,Tao:2012_space_charge} 
which is akin to the Child-Langmuir current limit for beams generated under the 
steady-state conditions\cite{Zhang:2016_review}. 
These short bunches are often generated by photoemission, and such bunches
inheret an initial profile similar
to that of the driving laser pulse profile.  Typically, the laser pulse has an in-plane, 
``transverse'' extent that is of order one hundred microns and a duration on the order
of fifty femtoseconds, and these parameters translate into an initial electron bunch with similar 
transverse extents and sub-micron widths\cite{King:2005_review}. 
After photoemission, the electrons are 
extracted longitudinally using either a DC or AC field typically 
in the 1-10 MV/m\cite{Srinivasan:2003_UED,Ruan:2009_nanocrystallography,van_Oudheusden:2010_rf_compression_experiment,Sciaini:2011_review} 
through tens of MV/m\cite{Musumeci:2010_single_shot,Weathersby:2015_slac,Murooka:2011_TED} ranges, respectively. However, the theoretical treatments of such 
``pancake-like'' electron bunch evolution have largely focused on the longitudinal 
dimension\cite{Luiten:2004_uniform_ellipsoidal,Siwick:2002_mean_field,Qian:2002_fluid_flow,Reed:2006_short_pulse_theory,Collin:2005_broadening}, 
and the few studies looking at transverse dynamics have either assumed a uniform-transverse distribution\cite{Collin:2005_broadening} or have 
looked at the effect of a smooth Gaussian-to-uniform evolution of the transverse profile on the evolution of the pulse in the longitudinal 
direction\cite{Reed:2006_short_pulse_theory,Portman:2013_computational_characterization}. Of specific note, only one analytic study found any indication, 
a weak longitudinal signal, of a shock\cite{Reed:2006_short_pulse_theory}.

\begin{figure*}
  \centering
  \begin{tabular}{ccc}
    \subfloat[full z-x]{\includegraphics[width=0.2\textwidth]{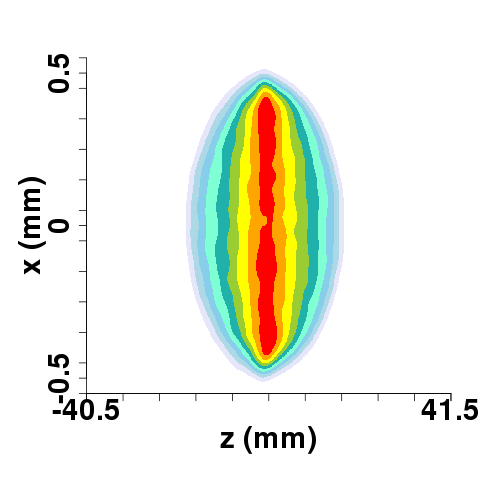}}&
    \subfloat[full x-y]{\includegraphics[width=0.2\textwidth]{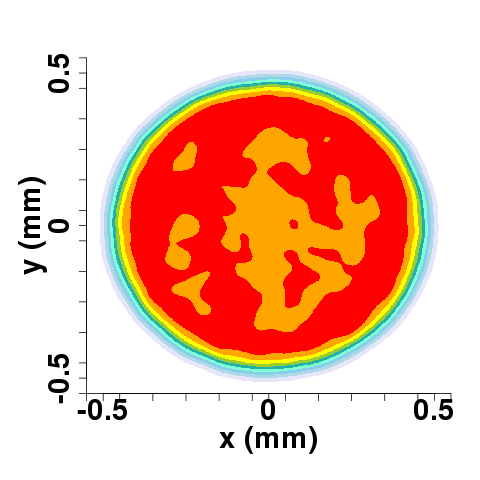}}&
   \\[-3mm]
    \subfloat[sliced z-x]{\includegraphics[width=0.2\textwidth]{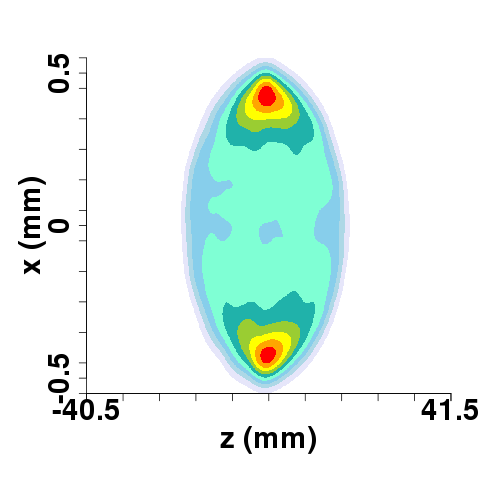}}&
    \subfloat[sliced x-y]{\includegraphics[width=0.2\textwidth]{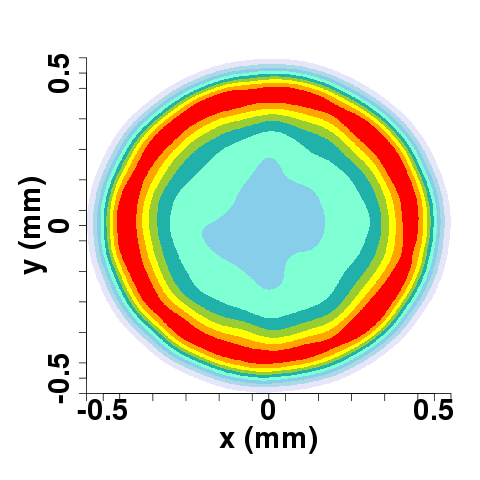}}&
     \multirow{2}{*}[8cm]{\subfloat[sliced radial evolution]{\includegraphics[width=0.6\textwidth]{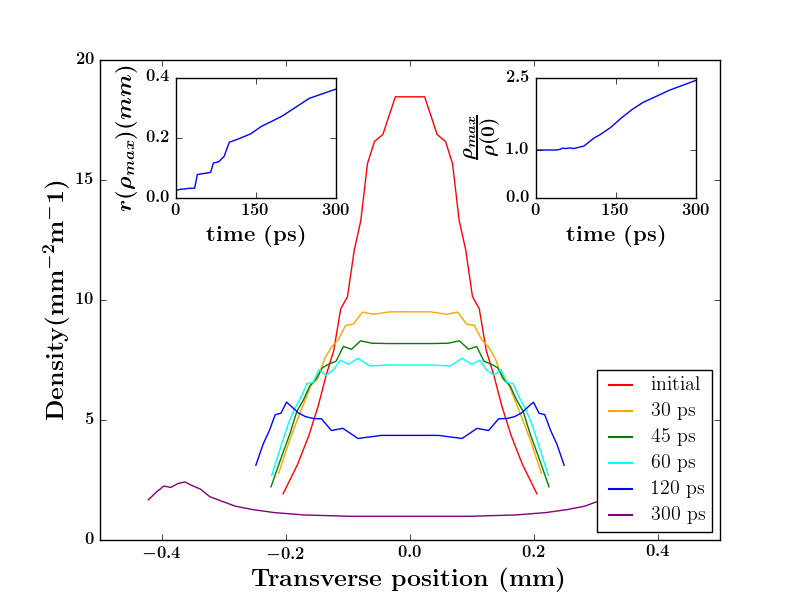}}}\\
  \end{tabular}
\caption{\label{fig:distribution substructure} (a-d.) Two dimensional projections of $1 \times 10^6$ electron positions simulated 
with the N-particle code, COSY, for approximately 300 ps after injection with a Gaussian ($\sigma_r = 100 ~\mu m$) 
transverse profile into a cavity with an electric field of 10 MV/m. Colors from white to red indicate electron density on a linear scale. 
 (a.) and (b.) are projections of the full distribution to the x-z and x-y planes, respectively. 
(c.) and (d.) are x-z and x-y projections, respectively, of just the portion of the distribution within 20\% of the standard deviation of the median value of y and z, 
respectively. Notice the ring-like substructure that is evident in the ``slices", (c.) and (d.) but absent from the full distribution projections, 
(a.) and (b.). (e.) N-particle radial-distributions obtained near the longitudinal median plane of the bunch at various times. 
Density is calculated by binning 1000 macroparticles and assigning the resulting density to the average radial position of those particles. 
The initial distribution is sampled from a Gaussian, and the square-like nature of the plot results from the discreteness of the bins. 
The sub-graph in the upper left corner shows the position of maximum density as a function of time, which is non-zero at initial time due to binning resulting 
in a non-zero minimum radial position. The sub-graph in the upper right shows the ratio of the maximum density to the density at the minimum $r$ value. 
Notice, the ``phase transition" in the 45-80 ps range where the location of the non-zero, non-stochastic peak first appears well away from the origin.
}
\end{figure*} 

On the other hand, an attractive theoretical observation is that an ellipsoidal cloud of cool, 
uniformly distributed charged particles has a linear electric field within the ellipsoid which results in maintenance of the uniform charge density as 
the cloud spreads \cite{Grech:2011_coulomb_explosion}. In the accelerator community, such a uniform distribution is a prerequisite in
employing techniques such as emittance compensation\cite{Rosenzweig:2006_emittance_compensation} as well as forming the basis of other theoretical analyses. 
It has long been proposed that such a uniform ellipsoid may be generated through 
proper control of the transverse profile of a short charged-particle bunch emitted from a source into vacuum\cite{Luiten:2004_uniform_ellipsoidal}, 
and experimental results have shown that an electron cloud emitted from a photocathode and rapidly 
accelerated into the highly-relativistic regime can develop into a final ellipsoidal profile 
characteristic of a uniform charge distribution\cite{Musucemi:2008_generate_uniform_ellipsoid}. Contrary to expectations from the free expansion 
work but consistent with the longitudinal analyses, this shadow lacks any indication of a peripheral region of high-density shocks. 
However, recent work has indicated that a substantial high density region may indeed form in the transverse direction\cite{Williams:2017_transverse_emittance}, 
and N-particle simulation results, as demonstrated in Fig. (\ref{fig:distribution substructure}), 
demonstrate a rapidly-developed substantial ring-like shock circumscribing the median of the bunch
when the bunch starts from sufficient density.
Moreover, this shock corresponds to a region of exceedingly low brightness, or conversely, high, local
temperature, and that experiments show that removal of this region results in a dramatic 
increase in the bunch brightness\cite{Williams:2017_transverse_emittance},
which we term ``Coulomb cooling'' as it is similar to evaporative cooling in the fact that
the ``hottest'' charged particles are removed from the distribution's edge thus leaving
behind a higher-quality, cooler bunch.  

To understand Coulomb cooling, we first investigate this transverse shock.
Here we demonstrate the formation of a ring-like shock within N-particle simulations
\cite{Berz:1987_cosy,Zhang:2015_fmm_cosy} 
of electron bunches with initial transverse Gaussian profile and offer an explanation of why this phenomena has not been noted previously within the UED literature. We then utilize a Poisson fluid approach to derive analytic predictions for the expansion dynamics in planar (1D), 
cylindrical, and spherical geometries, and we derive conditions for the emergence 
of density peaks distinct from any initial density maximum. We show that peak formation has a strong dependence on dimension, with one dimensional systems 
less likely to form shocks, while in cylindrical and spherical geometries bunching is more typical. Particle-in-cell (PIC) methods, utilizing 
WARP\cite{Friedman:2014_warp}, and 
N-particle simulation are then used to validate the analytical predictions for peak emergence.  

\section{Observation of Transverse Shock}
One reason that a transverse shock has not been seen previously in N-particle simulations
is apparent in Fig. (\ref{fig:distribution substructure}). We consider pancake electron bunches typical of
 100keV ultrafast electron microscopy, and we consider the thin direction of the bunch to be the z-axis. Previous studies of the 
expansion dynamics of these bunches, including our own work,
have looked at the projection of the particle density distribution to the x-z plane\cite{Luiten:2004_uniform_ellipsoidal,Musucemi:2008_generate_uniform_ellipsoid,Portman:2014_image_charge,Morrison:2013_measurement,Li:2008_quasiellipsoidal}. 
Fig. (\ref{fig:distribution substructure})
shows that by projecting the distribution in this manner, and with the ability to statistically discern
density fluctuations at about the 10\% level, results in what appears to be a uniform distribution; however,
by restricting the projection to only electrons near the median of the bunch, a restriction that can only
be done computationally presently, results in evidence of a transverse ring-like density substructure near the median longitudinal (z)
position.

\begin{figure*}
  \centering
  \begin{tabular}{ccc}
    \subfloat[$N = 1,000$]{\includegraphics[width=0.3\textwidth]{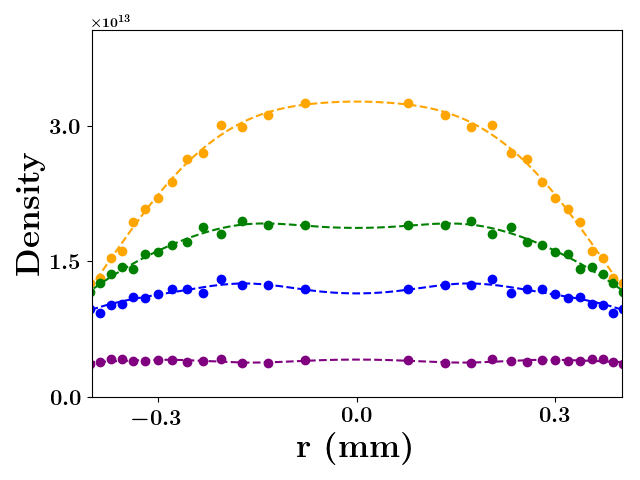}}&
    \subfloat[$N = 10,000$]{\includegraphics[width=0.3\textwidth]{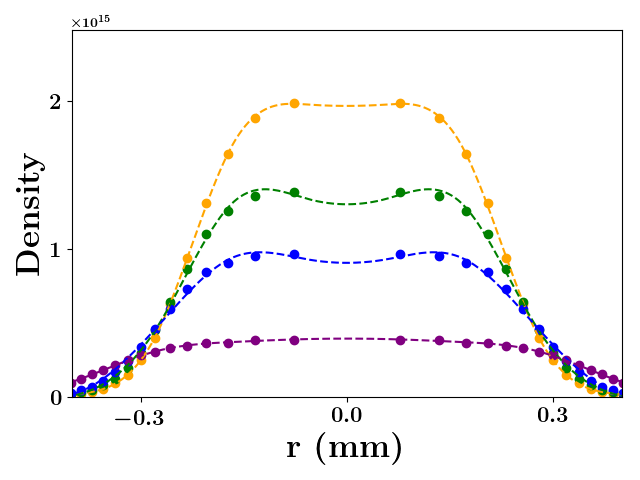}}&
    \subfloat[$N = 100,000$ ]{\includegraphics[width=0.3\textwidth]{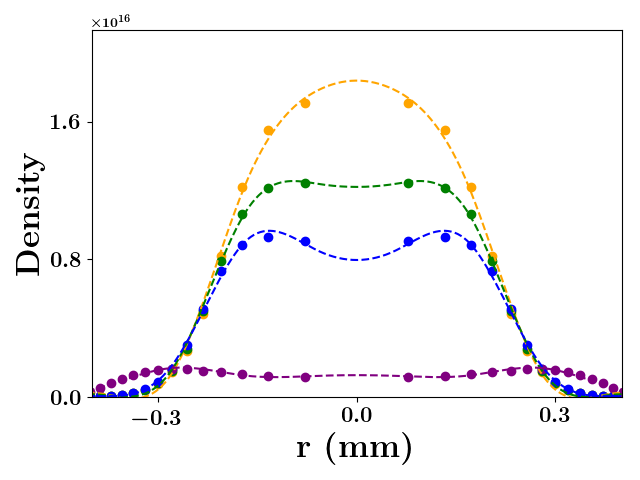}}
  \end{tabular}
\caption{\label{fig:average distribution} Average density near the z-median of 30 simulations calculated from 
bunch profiles evaluated at different times: $5  ~ \tau_p$ (yellow), $6  ~ \tau_p$ (green), 
$7  ~ \tau_p$ (blue), and $10 ~ \tau_p$ (purple) where 
$\tau_p = 2 \pi \sqrt{\frac{m \epsilon_0}{n e^2}}$, $n = \frac{N}{\pi \sigma_r^2 \sigma_z}$,
and $\sigma_z \approx 0.4~\mu$m, for different values of the total number of electrons, $N$.  
Dotted lines represent spline fits of 
order 3 with 10 knots.    }
\end{figure*} 

\begin{figure}
  \centering
   \includegraphics[width=0.45\textwidth]{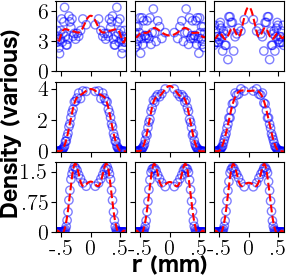}
\caption{\label{fig:splined_grid} Density near the z-median of simulated pancake bunches with 
transverse Gaussian profiles ($\sigma_r = 100 ~ \mu$m)  in an extraction field of 
10 MV/m.  Each figure is the transverse radial density of a section of width $\sigma_z \approx 0.4~\mu$m
for different initial conditions and different numbers of electrons, at time 
$10 ~ \tau_p$ where 
$\tau_p = 2 \pi \sqrt{\frac{m \epsilon_0}{n e^2}}$ is the plasma frequency; where $n = \frac{N}{\pi \sigma_r^2 \sigma_z}$.  
The number of electrons in each horizontal panel is different and 
equal to $N = 1,000$ (top), $N=10,000$ (middle), and $N = 100,000$ (bottom).   
For the density at $10 ~ \tau_p$, 30 cylindrical shells of equal volume and 
length $\sigma_z$ partitioned the distribution
out to $0.6$ mm, and the numbers of electrons in each of these shells were used to 
calculate a density at the shell's average radius. Due partially to the different numbers of electrons
and partially due to the fact that the longer simulations, namely the simulation with $N = 1000$, resulted
in significantly more electrons migrating out of the analysis region as a result of the initial 
velocity spread ,
the density scales are different for the three rows in the figure:
$\frac{1}{(0.1 mm)^3}$ for the top row, 
$\frac{0.1}{(0.01 mm)^3}$ for the middle row, 
and $\frac{1}{(0.01 mm)^3}$ for the bottom row.
Red dashed lines represent splines of 
order 3 with 10 knots.  Notice the clear presence of a shock  for the case $N=100,000$, an ambiguous
shock at $N=10,000$, and essentially noise at $N=1,000$.}
\end{figure}

\begin{figure}
  \centering
  \begin{tabular}{cc}
    \includegraphics[width=0.45\textwidth]{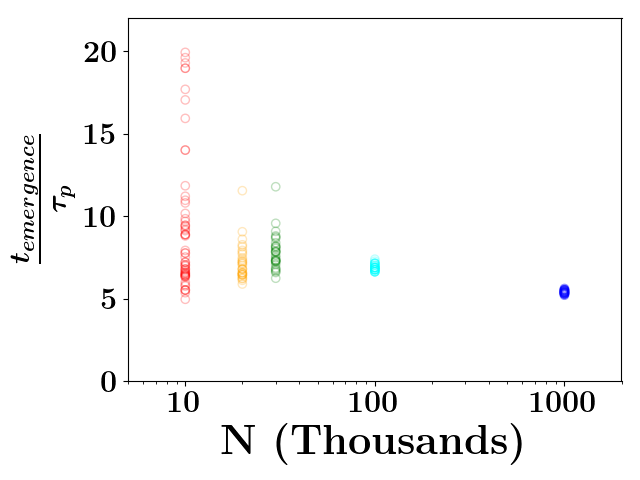}
  \end{tabular}
\caption{\label{fig:emergence time} The emergence time divided by the plasma
period as a function of the number of electrons in the initial Gaussian profile with 
$\sigma_r \approx 100 ~\mu$m and $\sigma_z \approx 0.4 \mu$m.  Emergence time was determined
as the first time the density away from the inner-most-value exceeded the inner-most-value by 2\%.  
Notice that the emergence time converges to about $5~\tau_p$ for high densities, but at low 
densities the emergence time has high variability with a median shifted to higher multiples of the plasma period.  }
\end{figure} 
To better understand when this shock emerges, simulations with the same 
distribution parameters ($\sigma_r \approx 100 ~\mu$m and $\Delta z \approx 0.4~\mu$m) but
various numbers of electrons were run.  The average radial density was calculated for 30
instances of bunches with 1 thousand, 10 thousand, and 100 thousand electrons.  As can be seen in Fig. (\ref{fig:average distribution}),
the shock emergence is present for bunches with 100 thousand electrons but not for those with 1 thousand electrons. 
  The case of bunches with 10 thousand electrons suggests the emergence of 
the shock, but the shock becomes less defined at later times.  Fig (\ref{fig:splined_grid}) shows nine density profiles  
at time $10 ~ \tau_p$, where $\tau_p$ represents the plasma period, with $\tau_p = 2 \pi \sqrt{\frac{m \epsilon_0}{n e^2}}$ and
$n = \frac{N}{\pi \sigma_r^2 \sigma_z}$.
The dots in each figure indicate average densities in cylindrical rings, originating from three randomly chosen initial conditions (figures in each row)  
  and for three values of the total number of electrons $N$: 1 thousand, 10 thousand, and 100 thousand electrons; for the top, middle, and bottom rows, respectively.
 These representative density plots support the conclusion that the shock is only present in the 
case of bunches with 100 thousand electrons; where the spline fit to the data indicates a significant peak
removed from the center of the bunch in all instances examined.  
As expected, the density of bunches with one thousand electrons is noisy due to low statistics
both from the small number of electrons in the simulation and the large proportion of electrons
that spread beyond the analysis region due to the initial velocity spread; and the density profile of 
bunches with 10 thousand electron has a consistent general
shape that fits well to the spline fit but lacks significant emergent peaks; which are
the indicators of shock formation. 

We define the emergence time as the time at which peaks indicative of a shock 
emerge in the dynamics of Coulomb clouds.  Fig. (\ref{fig:emergence time})
shows the dependence of the emergence time, and its variability, on the number 
of electrons in a bunch.  It also shows very clearly that the emergence time is 
proportional to the plasma period, $\tau_p$.  As can be seen in this figure, the spread in the emergence time
 is large for a bunch with 10 thousand electrons, but this spread decreases as the density of the bunch increases.  
For bunches with $N\ge 100,000$ the spread in the emergence time is
small, moreover the emergence time appears to converge toward approximately $5~\tau_p$ at large $N$ (for Gaussian initial distributions). 
We note here that for Gaussian pulses with similar spatial and temporal extents, simulations at and above $10$ 
million electrons, a goal of the community\cite{BES_report:2016_electron_sources}, result
in relativistic velocities as a result of the stronger space-charge effects.  
As the discussion here focuses on non-relativistic physics, we present data for up to
 1 million electrons, where the velocity obtained from the self-field 
remains non-relativistic.  

The results presented in Figs. 2-4 are a second reason that shock formation has not been seen previously in studies of electron bunches.  
Specifically, most  work has been conducted using $\le 10,000$ electrons with a transverse standard deviation of 100 $\mu$m, 
and  in the regime where there is no consistent emergence of a shock. 
Moreover, the fact that the 
non-relativistic evolution of the bunch profile has a time scale proportional to the
plasma period,
a fact that we derive  under special 
geometries later in this manuscript, means
that higher density bunches result in faster, more consistent evolution of the 
transverse profile.  In other words, the
emergence time of a shock happens earlier as the density
of the bunch is increased.  Specifically, a transverse shock emerges at on the order of 
$50$ ps for an initially Gaussian profile ($\sigma_r = 100 ~\mu$m with sub-micron length)
with $10^6$ electrons, which is the number of electrons which is the current goal
for the diffraction community\cite{BES_report:2016_electron_sources}.  
This implies that for modern bunches, this transverse shock is
happening well within the photoemission gun before the onset of the relativistic regime.  The
goal of $10^8$ electrons for the imaging community needs to be further examined as the 
transverse velocity spread will be relativistic, but we expect to find this effect there as well, however 
we expect that it occurs at short times, of order a few picoseconds.

\section{1D model}

As noted in the introduction, formation of a shock in the longitudinal direction 
of an expanding pancake pulse has not been observed, and the analysis of 
Reed~\cite{Reed:2006_short_pulse_theory} demonstrates that this is true 
for cold initial conditions.   Here we re-derive this result using an elementary 
method, which enables extension to include the 
possibility of an initial chirp; and we find chirp conditions at which shock formation 
in the longitudinal direction can occur. 

Consider the non-relativistic spreading of an electron bunch in a one dimensional model, which is a good early time approximation 
to the longitudinal spreading of a pancake-shaped electron cloud generated at a photocathode. In one dimensional models, 
the  density, $\rho$, only depends on one coordinate, which we take to be $z$. We also take $\rho$ to be normalized so 
that its integral is one. For the sake of readability, 
denote the position of a particle from the Lagrangian perspective to be 
$z = z(t)$ and $z_0 = z(0)$. The acceleration of a Lagrangian particle is
\begin{align}
  a(z;t) = \frac{qQ_\text{tot}}{2 m \epsilon_0} \delta \sigma
\end{align}
where $q$ is the charge of the particle (e.g. electron), $m$ is its mass, $Q_{\text{tot}}$ is the total 
charge in the bunch, and 
\begin{align}
  \delta \sigma 
                       = \int_{-z}^{z} \rho(\tilde{z};t) d\tilde{z}
\end{align} 
The key observation enabling analytic analysis 
is that if the flow of electrons is lamellar, so that there is no crossing of particle trajectories,
then  these integrals and the acceleration calculated from them are
time independent and hence may be determined from the initial distribution. Therefore, we denote 
$a(z;t) = a_0$, $\rho_0 = \rho(z;0)$, and $\delta \sigma =  \int_{-z_0}^{z_0} \rho_0(\tilde{z})d\tilde{z}$. 
Moreover, due to the fact that for any particle trajectory, the acceleration is constant and given by $a_0$, the 
Lagrangian particle dynamics reduces to the elementary constant acceleration kinematic equation
\begin{equation}\label{eq:non-relativistic position}
  z(t) = z_0 + v_0 t + {1\over 2} a_0 t^2
\end{equation}
where $v_0$ is the initial velocity of the charged particle that has initial position $z_0$. 
Notice that both $v_0$ and $a_0$ are functions of the initial position, $z_0$, and we shall see later that the derivatives of
these parameters, $v' = \frac{dv_0}{dz_0}$  and $a' = \frac{da_0}{dz_0}$ are important in describing 
the relative dynamics of Langrangian particles starting at different initial positions. Moreover, the special case of 
$v_0 = 0$ everywhere, which we will call the cold-case, is commonly assumed in the literature, and we now examine this case in detail.

First we consider the speading charge distribution within the Eulerian perspective, where $z$ is an independent variable 
instead of it describing the trajectory of a particle. We denote the charge distribution at all times to be 
$Q_\text{tot} \rho(z; t)$ with $\rho(z; t)$ a unitless, probability-like density and $Q_\text{tot}$ the total charge per unit area
 in the bunch. 
Since particle number is conserved, we have
\begin{align}
  \rho(z; t)dz = \rho_0 dz_0 
\end{align}
so that in the non-relativistic case derived above
\begin{align}
  \rho(z,t) &=  \rho_0 \left({dz\over dz_0}\right)^{-1} =\frac{ \rho_0}{1 + v_0' t +\frac{1}{2} a_0' t^2} \label{eq:1D density evolution}
\end{align}
Notice  that the derivative of the acceleration with respect to the initial position is directly proportional to the initial distribution, so that for initial distributions that are symmetric about the origin,
\begin{align}\label{eq:first order acceleration}
  a_0' &= \frac{q Q_\text{tot}}{2 m \epsilon_0}  \frac{d{\delta \sigma}}{dz_0}\nonumber\\
          &= \frac{q Q_\text{tot}}{m \epsilon_0}\rho_0
\end{align}
 Plugging Eq. (\ref{eq:first order acceleration}) into Eq. (\ref{eq:1D density evolution}), 
we get
\begin{align}
  \rho(z,t) &=  { \rho_0 \over  1 + v_0' t +  \frac{q Q_\text{tot}}{2 m \epsilon_0} \rho_0t^2}\label{eq:1D density evolution subbed}\\
  \frac{d}{dz} \rho(z,t) &= \frac{ \rho_0'\left( 1  + v_0' t\right) - \rho_0v_0''t }{\left(1 + v_0' t +  \frac{q Q_\text{tot}}{2 m \epsilon_0} \rho_0t^2\right)^3}\label{eq:1D rho derivative} 
\end{align}
where $\rho_0' = \frac{d\rho_0}{dz_0}$ and $v_0'' = \frac{d^2v_0}{dz_0^2}$.  A detailed derivation of the second expression is in Appendix 
\ref{ap:1D appendix}.

For the cold-case, Eq. (\ref{eq:1D density evolution subbed})  
reduces to the density evolution equation derived by Reed\cite{Reed:2006_short_pulse_theory} using different methods. Also, in the cold-case, 
Eq. (\ref{eq:1D rho derivative}) simplifies into a proportionality between the initial slope of the distribution and the slope of the distribution at any later time. 
Therefore, a charge distribution that is initially at rest and unimodal, i.e only a single initial location has non-zero density with $\rho_0' = 0$, 
never develops a dynamically generated second maximum. This
explains why we should not expect to see an emergent shock in the longitudinal direction; provided the 
1D model is applicable and cold initial conditions are valid. 

However, if particles in the initial state have an initial velocity that depends on initial position, i.e. $v_0(z_0)$, 
then density peaks will emerge at $z$ when $t = \frac{1}{v_0''\rho_0 - v_0' \rho_0'}$.  This occurs 
 at positive time if $v_0'' \rho_0 > v_0' \rho_0'$.  In the special case 
$v_0'' \rho_0 = v_0' \rho_0'$, the distribution may be reframed as a cold-case distribution starting from
$t = -\frac{m c_1 \epsilon_0}{q Q_\text{tot}}$ for some $z_0$-independent constant $c_1$ with velocity units when
$\frac{m c_1^2 \epsilon_0}{q Q_\text{tot}} < 1$ or 
a distribution starting from a singularity with velocity distribution 
${\tilde{v}}_0 = c_1 \left( \rho_0 - \frac{1}{2} \delta\sigma\right)$. 
As noted earlier, the function 
$a_0(z_0)$ is monotonically increasing as a function of distance from the center of the pulse, which means that electrons at the edges of the bunch 
always have larger accelerations away from the center of the pulse than electrons nearer the pulse center. Thus crossover, where an inner electron moves past an 
outer electron, cannot occur unless the initial velocities of inner electrons overcome this relative acceleration.

A practical case where crossover may be designed is where the initial distribution has an initial velocity chirp, 
i.e. $v_0 = c z_0$ where $c$ has units of inverse time.   Intuitively we can expect that the velocity chirp needs to 
be negative in order for crossover to occur. 
To find the crossover time, we consider the time at which two electrons that were initially apart,  are at the same position at the same time.
In this case it is straight forward to find 
the time at which crossover occurs by considering an electron at initial position $z_0$, and a second electron at position 
$z_0 + \delta z_0$.  
Before either of these electrons experiences a crossover Eq.  (\ref{eq:non-relativistic position}) is valid, and setting 
$z(z_0, t_x) = z(z_0 + dz_0, t_x)$
reduces to,
\begin{align}
  At^2_x + Bt_x + 1 = 0
\end{align}
where $A = \frac{q}{2m \epsilon_0} \rho_0(z_0)$ and $B = \frac{dz}{dz_0}$.  Solving the quadratic equation leads to the crossover time given by
\begin{align}
  t_x &= \frac{-B \pm \sqrt{B^2 - 4A}}{2A}
\end{align}
Since $A$ is always positive, the square root is real only if $B^2$ is larger than $4A$. Moreover
the time is only positive if $B$ is negative. Therefore crossover only occurs if the chirp 
has a negative slope, as expected on physical grounds. The conditions for tuning the chirp to 
produce crossover in 1D are then
\begin{align}
  \frac{dv_0}{dz_0} &< 0\label{eq:negative chirp}\\
  \left| \frac{dv_0}{dz_0}\right| &\ge \sqrt{\frac{2q}{m\epsilon_0}\rho_0(z_0)}\label{eq:sufficient chirp})
\end{align}

The results above are applicable to the spreading in the longitudinal direction 
 of non-relativistic pancake bunches, because the expression Eq. (\ref{eq:non-relativistic position}) is linear in acceleration. 
In that case,  the position of a charged particle at 
any time can be calculated from a superposition of the contribution from the space-charge field and any external constant field such as a constant and uniform extraction field. 
In that case, the space charge field leads to spreading of the pulse, while the extraction field leads solely to an acceleration of the center of mass of the entire bunch.  
In that case the center of mass and spreading dynamics are independent and can be decoupled.
The extension of the description above to asymmetric charge density functions is also straightforward, as is the inclusion of an image field at the photocathode. 
Moreover, inclusion of these effects does not change Eq. (\ref{eq:1D density evolution subbed}), Eq. (\ref{eq:1D rho derivative}), nor the conclusions we have drawn from them.
These results apply generally to all times before the initial crossover event within the evolution of the bunch, and once 
crossover occurs, the distribution can be reset with a new Eq. (\ref{eq:1D density evolution subbed})  to follow further density evolution.

As we show in the next section the one-dimensional results do not apply, even qualitatively, to 
 higher dimensions, as the constant acceleration situation is not valid and crossover can 
occur even with cold initial conditions, as demonstrated in the simulations presented in previous 
section.  In the next section we present fluid models in higher dimensions where the origin of 
these new effects is evident. 

\section{Cylindrical and Spherical Models}

The methodology for the cylindrical and spherical systems is similar so we 
develop the analysis concurrently.   

Consider a non-relativistic evolving distribution 
$Q_\text{tot} \rho(r,t)$, where $\rho(r,t)$ is again taken to be the unitless
particle distribution and $Q_\text{tot}$ is again the total charge in the bunch.    In a system with cylindrical symmetry,
the mean field equation of motion for a charge at $r \equiv r(t)$ is given by
\begin{equation}\label{eq:2D second derivative}
\frac{d \vec{p}_\text{r}}{dt} = { q Q_\text{tot} \lambda (r,t)\over 2 \pi \epsilon_0 r}\hat{r}, 
\end{equation}
where $\vec{p}_\text{r}$ is the momentum of a Lagrangian particle, $\lambda(r,t)$ is the cummulative 
distribution function (cdf)
\begin{equation}
\lambda(r,t) = \int_0^{r} 2\pi \tilde{r} \rho(\tilde{r},t) d\tilde{r}
\end{equation}
 and $Q_\text{tot} \lambda(r;t)$ is the charge inside radius $r$.
Analogously in a system with spherical symmetry,
the equation of motion for an electron at position $r$ is given by
\begin{equation}\label{eq:3D second derivative}
\frac{d \vec{p}_\text{r}}{dt} = { q Q_\text{tot} P(r,t)\over4\pi \epsilon_0 r^2}\hat{r}, 
\end{equation}
where $P(r,t) = \int_0^{r} 4\pi \tilde{r}^2 \rho(\tilde{r},t) d\tilde{r}$ is the cdf and  
$Q_\text{tot} P(r;t)$ is the charge within the spherical shell of radius $r$.  
Notice, $r$ in Eq. (\ref{eq:2D second derivative}) denotes the cylindrical radius while
$r$ in Eq.  (\ref{eq:3D second derivative}) represents the spherical radius.  
In both cases, before
any crossover occurs, the cdf of a Lagrangian particle is constant in time.  For simplicity we write
$\lambda(r,t) = \lambda(r_0,0) \equiv \lambda_0$
and $P(r,t) = P(r_0,0) \equiv P_0$ for a particle starting at $r_0 \equiv r(0)$.
In other words, since $Q_\text{tot} \lambda_0$ and
$Q_\text{tot} P_0$ can be interpreted as the charge contained in the appropriate 
Gaussian surface, if we
track the particle that starts at $r_0$, these contained charges should remain constant before
crossover occurs.
It is convenient to also define the average particle density to be
$\bar{\rho}_0 = \frac{\lambda_0}{\pi r_0^2 }$ in the cylindrically symmetric case and 
 $\bar{\rho}_0 = \frac{3 P_0}{4\pi r_0^3 }$ in the spherically symmetric case.  Notice that
 these average particle densities are a function solely of $r_0$, and
 we will use these parameters shortly.
Eq. (\ref{eq:2D second derivative}) and 
Eq. (\ref{eq:3D second derivative}) may now be rewritten as, for the cylindrical and spherical cases respectively
\begin{align}
  \frac{d p_r}{dt} &= { q Q_\text{tot} \lambda_0\over 2 \pi \epsilon_0 r}\label{eq:2D equation of motion}, \\ 
  \frac{d p_r}{dt} &= { q Q_\text{tot} P_0\over 4 \pi \epsilon_0 r^2}\label{eq:3D equation of motion}, 
\end{align}
which apply for the period of time before particle crossover.  Note that unlike the one dimensional case, 
 in two and three dimensional systems the acceleration on a Lagrangian particle 
is not constant, and has a  time dependence through the time dependent position 
$r=r(r_0:t)$ term in the denominator.

 Since Eq. (\ref{eq:2D equation of motion}) and Eq. (\ref{eq:3D equation of motion}) 
 represent the force on the particle in the cylindrical and spherical contexts, respectively, 
 we can integrate over the particle's 
 trajectory to calculate the change in the particle's energy. 
 Integrating from $r_0, 0$ to $r, t$ gives for the cylindrical and spherical cases respectively
\begin{align}
  E(r,t) - E(r_0,0) &= \frac{q Q_\text{tot} \lambda_0}{2 \pi \epsilon_0} ln\left(\frac{r}{r_0}\right)\label{eq:2D energy}\\
  E(r,t) - E(r_0,0) &= \frac{q Q_\text{tot} P_0}{4 \pi \epsilon_0} \left(\frac{1}{r_0} - \frac{1}{r}\right)\label{eq:3D energy}
\end{align}
where the term on the right side of the equality can be interpreted as  the change in the potential energy within the self-field
of the bunch.  

These expressions are fully relativistic, and in the non-relativistic limit, we can derive implicit position-time relations for the particle
by setting the energy difference equal to the non-relativisitic kinetic energy $mv^2/2$, and integrating.  
The details of this derivation have been placed in Appendix \ref{ap:time-location appendix}, and 
the resulting expressions in the cold-case for the cylindrical and spherical systems are respectively
\begin{align}
  t &= \frac{\bar{\tau}_{p,0}}{\pi} \frac{r}{r_0} F\left(\sqrt{ln\left(\frac{r}{r_0}\right)}\right)\label{eq:2D time}\\
 t &= \sqrt{\frac{3}{2}}\frac{\bar{\tau}_{p,0}}{2\pi} \left( \tanh^{-1} \left( \sqrt{1 -  \frac{r_0}{r}} \right) + \frac{r}{r_0}\sqrt{1 -  \frac{r_0}{r}}\right)\label{eq:3D time}
\end{align}
where $F(\cdot)$ represents the Dawson function  and $\bar{\tau}_{p,0}$ represents the plasma period 
determined from the initial conditions: 
$\bar{\tau}_{p,0} = 2\pi \sqrt{\frac{m\epsilon_0}{q Q_\text{tot} \bar{\rho}_0}} = \frac{2\pi}{{\bar{\omega}}_0}$, 
indicating that the
appropriate time scale is the scaled plasma period as seen in Figs. (\ref{fig:average distribution}) and (\ref{fig:emergence time})
for the case of pancake bunches used in ultrafast electron diffraction systems.  Eq. (\ref{eq:3D time}) and its derivation is equivalent
to previous time-position relations reported in the literature\cite{Boyer:1989_kinetic_energy,Last:1997_analytic_coulomb_explosion} although the previous work
did not identify the plasma period as the key time-scale of Coulomb spreading  processes and  
cylindrical symmetry was not discussed (Eq. (\ref{eq:2D time})).

The time-position relations detailed in the equations above depend solely on the amount of charge 
nearer to the origin than the point in question, i.e. $Q_\text{tot} \bar{\rho}_0$, and not on the details of the distribution.  Notice however, that it is the 
difference between the time-position relationships of different locations where the 
details of the distribution become important and may cause neighboring particles to 
have interesting relative dynamics; leading to the possibility of shock formation in the density.

To translate the Lagrangian particle evolution equations above to 
an understanding of the  dynamics of the charge density distribution, we generalize Eq. (\ref{eq:1D density evolution}) to
\begin{align}
  \rho(r,t) = \rho_0 \left( \left(\frac{r}{r_0}\right)^{d-1} {\frac{dr}{dr_0}}\right)^{-1}\label{eq:density evolution general}
\end{align}
where d is the dimensionality of the problem, i.e. 1 (planar symmetry), 
2 (cylindrical symmetry), or 3 (spherical symmetry).
The factor in the denominator, ${\frac{dr}{dr_0}}$, may be 
determined implicitly from the time-position relations above, and the details are presented in 
Appendix \ref{ap:spatial derivatives}. 
The resulting expressions for the density dynamics,  in the cold case, for $d=2$ (cylindrical) and $d=3$ (spherical) cases are 
\begin{align}
  \frac{d r}{dr_0} &= \frac{r}{r_0} \left(1 + D_d(r_0) f_\text{d}\left(\frac{r}{r_0}\right)\right)\label{eq:dr over dr_0 no v_0}
\end{align}
where 
\begin{equation}
D_d = D_d(r_0) = \frac{d}{2} \left(\frac{\rho_0}{\bar{\rho}_0} - 1\right),
\end{equation}
is a function only of the initial position.  The functions $f_d$ for cylindrical systems is given by
\begin{equation}
f_\text{2}\left(\frac{r}{r_0}\right) = 2  \sqrt{\ln\left(\frac{r}{r_0}\right)}F\left(\sqrt{\ln\left(\frac{r}{r_0}\right)}\right),
\end{equation}
while for systems with spherical symmetry we find
\begin{equation}
f_\text{3}\left(\frac{r}{r_0}\right) = \frac{r_0}{r} \sqrt{1 - \frac{r_0}{r}} \tanh^{-1}(\sqrt{1 - \frac{r_0}{r}}) + 1 - \frac{r_0}{r}.
\end{equation}
Note that these are functions of the ratio $r/r_0$. 
The functions $f_d$ can also be written as mixed functions of $r$ and $t$, specifically
$f_2\left(\frac{r}{r_0}\right) = \frac{r_0}{r}\sqrt{\ln\left(\frac{r}{r_0}\right)}{\bar{\omega}}_0t$ and
$f_3\left(\frac{r}{r_0}\right) = \frac{r_0}{r}\sqrt{1 - \frac{r_0}{r}}{\bar{\omega}}_0t$.
However, care must be used when 
using these mixed forms as $r$ is implicitly dependent on $t$.  Here we work with
these functions in terms of relative position, $\frac{r}{r_0}$. 
Substituting Eq. (\ref{eq:dr over dr_0 no v_0}) into Eq. (\ref{eq:density evolution general}), 
we find that the density evolution in systems with cylindrical (d=2) and spherical symmetry(d=3)  can be compactly written 
as
\begin{align}
  \rho(r;t) &= \left(\frac{r_0}{r}\right)^d \frac{\rho_0}{1 + D_d(r_0)  f_\text{d}\left(\frac{r}{r_0}\right)}\label{eq:general density evolution}
\end{align}
This expression is general and can be applied to arbitrary, spherically symmetric or cylindrically symmetric initial conditions.  

Analogously to the 1D case the condition $\frac{dr}{dr_0} < 0$ results in particle crossover.  However, as 
detailed in Eq. (\ref{eq:dr over dr_0 no v_0}),  the sign of $\frac{dr}{dr_0}$ depends on the sign
of $1 + D_d(r_0) f_\text{d}\left(\frac{r}{r_0}\right)$.
It is very interesting to note that $ D_\text{d}(r_0)$ is
the deviation from a uniform distribution function, so that 
the $D$ functions are solely functions of the initial conditions and are positive at locations where the local density is larger than the average density at $r_0$, and negative when the local density
is smaller than the average density at $r_0$.   On the other hand, the functions $f_d$ are functions of the evolution of the 
Langrangian particle.
One immediate consequence of Eq.(\ref{eq:general density evolution}) is that for a uniform 
initial density distribution, for either  cylindrical or spherical systems, the 
corresponding $D$ function is zero at every location where the original density is defined.  Thus, the uniform density evolution
in Eq. (\ref{eq:density evolution general})
reduces to the generally recognized expressions:  $\rho(r,t) \pi r^2=  \rho_0 \pi r_0^2$ for the cylindrical case; and $\rho(r,t) \frac{4}{3} \pi r^3=  \rho_0 \frac{4}{3} \pi r_0^3$ in
the spherical case.  We provide additional details for the uniform distribution in the next section.
However, Eq. (\ref{eq:dr over dr_0 no v_0})
is general for any distribution before particle crossover, not just the uniform distribution.

\begin{figure}
  \centering
  \subfloat[$f_\text{2}\left(\frac{r(t)}{r(0)}\right)$]{\includegraphics[width=0.24\textwidth]{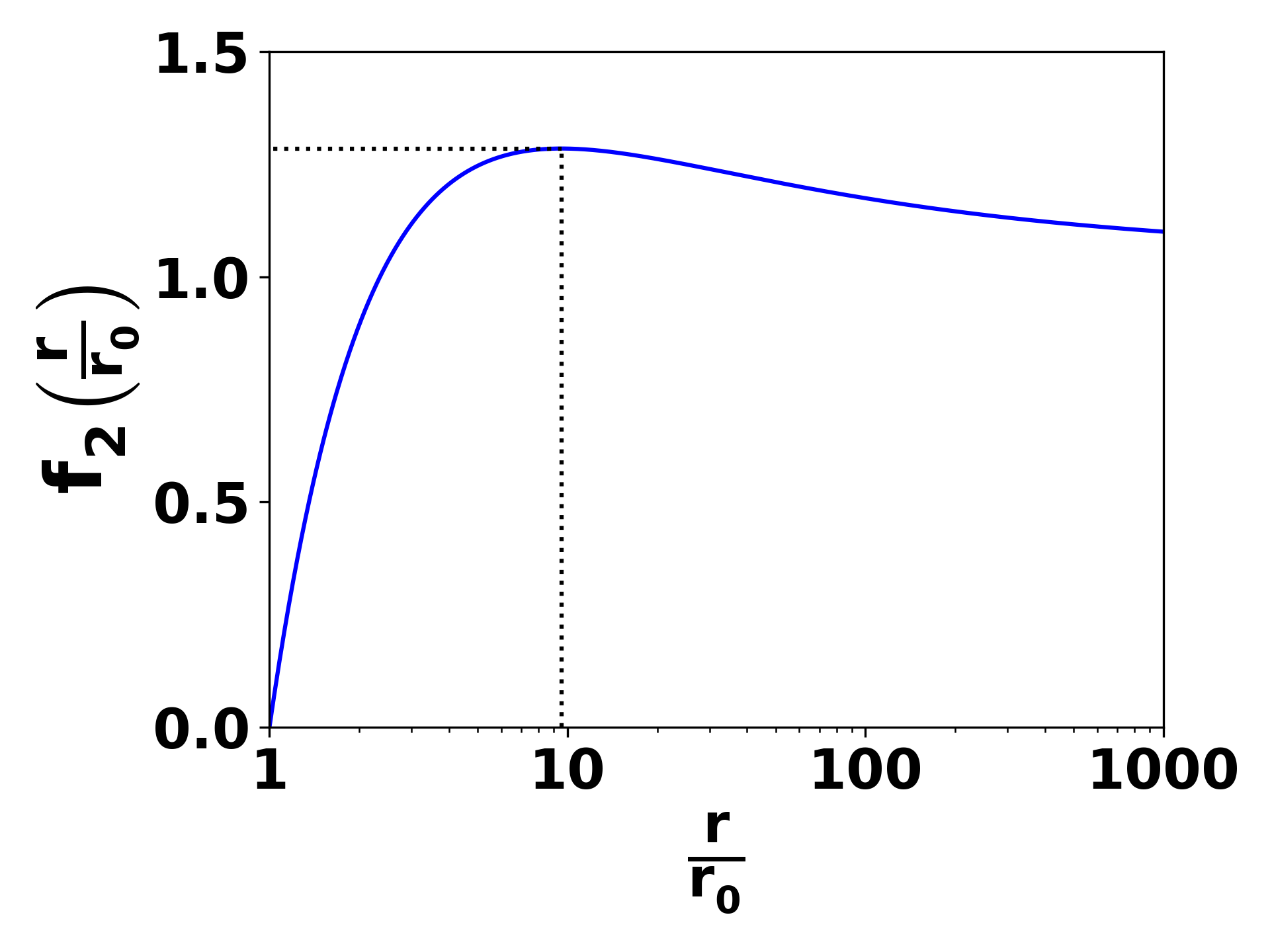}}
  \subfloat[$f_\text{3}\left(\frac{r(t)}{r(0)}\right)$]{\includegraphics[width=0.24\textwidth]{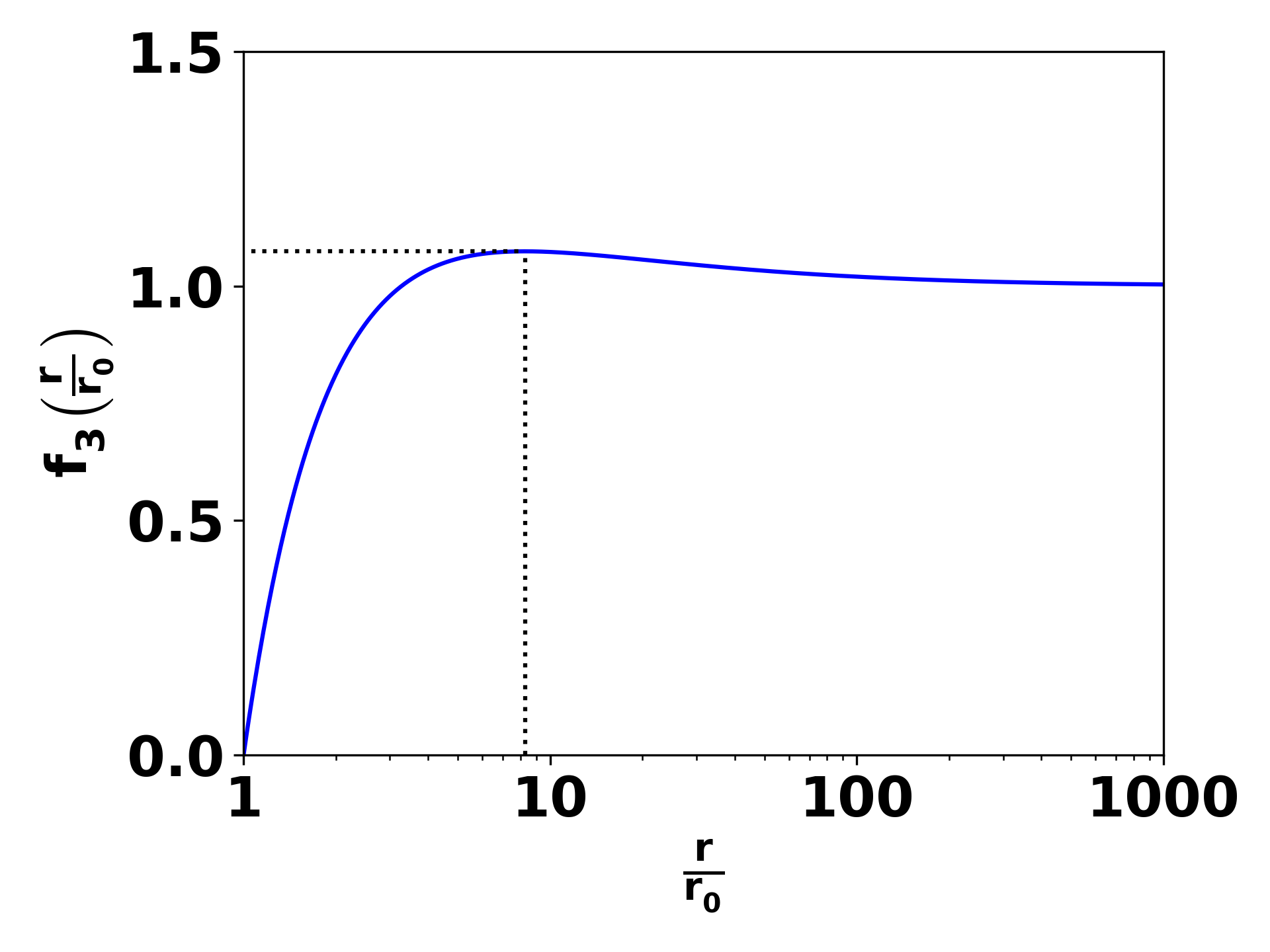}}\\
  \subfloat[$D_\text{2}(r(0))$ classification]{\includegraphics[width=0.24\textwidth]{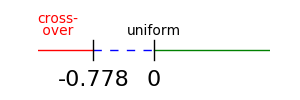}}
  \subfloat[$D_\text{3}(r(0))$ classification]{\includegraphics[width=0.24\textwidth]{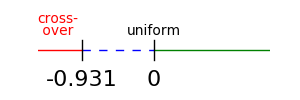}}
\caption{\label{fig:classifying D} (a-b.) Plot of the functions $f_\text{2}$ and $f_\text{3}$  against $\frac{r}{r_0}$.  
Since particles are always moving away from the origin, $\frac{r}{r_0} > 1$ for positive times and this ratio 
go to infinity for infinite time.  Both functions have a similar character with a maximum in the $8-10 ~ r_0$ range, and they both 
eventually approach one from above.  Dashed lines in the plots
indicate the location of the functions' maxima.  These functions allow
us to analytically classify the distribution by the initial value of 
$D_\text{d}(r_0) = \frac{d}{2} \left(\frac{\rho_0(r_0)}{\bar{\rho}_0(r_0)} - 1\right)$
as is done in (c-d.).  Specifically, crossover may occur at some time when $\frac{dr}{dr_0} < 0$. 
For negative values of $D_d$ crossover first occurs at the maximum of $f_d$ and the value of $D_d$ at 
which that occurs is given in the schematics of (c,d).   For values of $D_d$ that are more negative 
than this value, crossover will occurs at some time in some parts of the distrubiton.   For the uniform distribution, $D(r_0)=0$ for
all points inside the distribution.  The dashed blue line indicates expansion less quickly than 
the uniform distribution, while the green line indicates more rapid expansion than for the uniform
distribution.}
\end{figure} 
 
For a particle starting at position $r_0$ and having a deviation from uniform function $D_\text{d}(r_0)$, crossover occurs when the particle is at 
a position, $r$, that satisfies $f_\text{d}\left(\frac{r}{r_0}\right) = -1/D_d(r_0)$.  Since every particle moves toward positive $r$, every particle will have a time for which it will assume every value of the function $f\left(\frac{r}{r_0}\right))$.  The character of the two and three dimensional $f$'s are similar as can be seen in Fig (\ref{fig:classifying D}) where
the value of the function is plotted against $\frac{r}{r_0}$.
Specifically, both functions increase to a maximum and then asymptote towards 1 from above.  
This means that all density positions eventually experience uniform-like scaling since $\lim_{r \to \infty} f_{d}(\frac{r}{r_0}) = 1$
 results in Eq. (\ref{eq:density evolution general}) simplifying to $\rho \pi r^2 =  \rho_0 \frac{\pi r_0^2}{1 + D_\text{d}(r_0)}$ and 
$\rho \frac{4}{3} \pi r^3=  \rho_0 \frac{\frac{4}{3} \pi r_0^3}{1 + D_\text{d}(r_0)}$ in the cylindrical and spherical cases, respectively, for large enough $r$.
Notice, this  uniform-like scaling does not mean that the distribution goes to the uniform
distribution, which is what happens in 1D but need not happen under 
cylindrical and spherical geometries.

The main difference between the cylindrical and spherical symmetries are that the cylindrical function's maximum is larger than the spherical function's maximum; and we find $max(f_\text{2}) \approx 1.28$ 
while $max(f_\text{3}) \approx 1.07$.  Moreover the maximum of the cylindrical function occurs at a larger
value of $\frac{r_0}{r}$ than  that of the spherical function; specifically $r \approx 9.54 r_0$ instead of $r \approx 8.27 r_0$, respectively.  The first observation means cylindrical 
symmetry is more sensitive to the distribution than the spherical case, while the second observation  indicates that if crossover is going to occur for a specific particle, 
it will occur before the $r$ value for which the corresponding $f$ function is maximum (i.e. $r \approx 9.54 r_0$ or $r \approx 8.27 r_0$), 
otherwise the particle will never experience crossover.      
From this reasoning, we obtain the earliest time for crossover  by minimizing
 the time taken for a trajectory to reach the maximum of the function $f_d$,  with the crossover constraint $ {\frac{dr}{dr_0}} = 0$.
This may be achieved by using  Lagrange multipliers or by running calculations for a series of values of $r/r_0$ to find the 
position at which crossover happens first.    The mean field theory is valid before the minimum crossover time, and the results 
presented below are well below this time. 
\begin{figure*}
  \centering
  \begin{tabular}{cc}
    \subfloat[cylindrical symmetry uniform]{\includegraphics[width=0.45\textwidth]{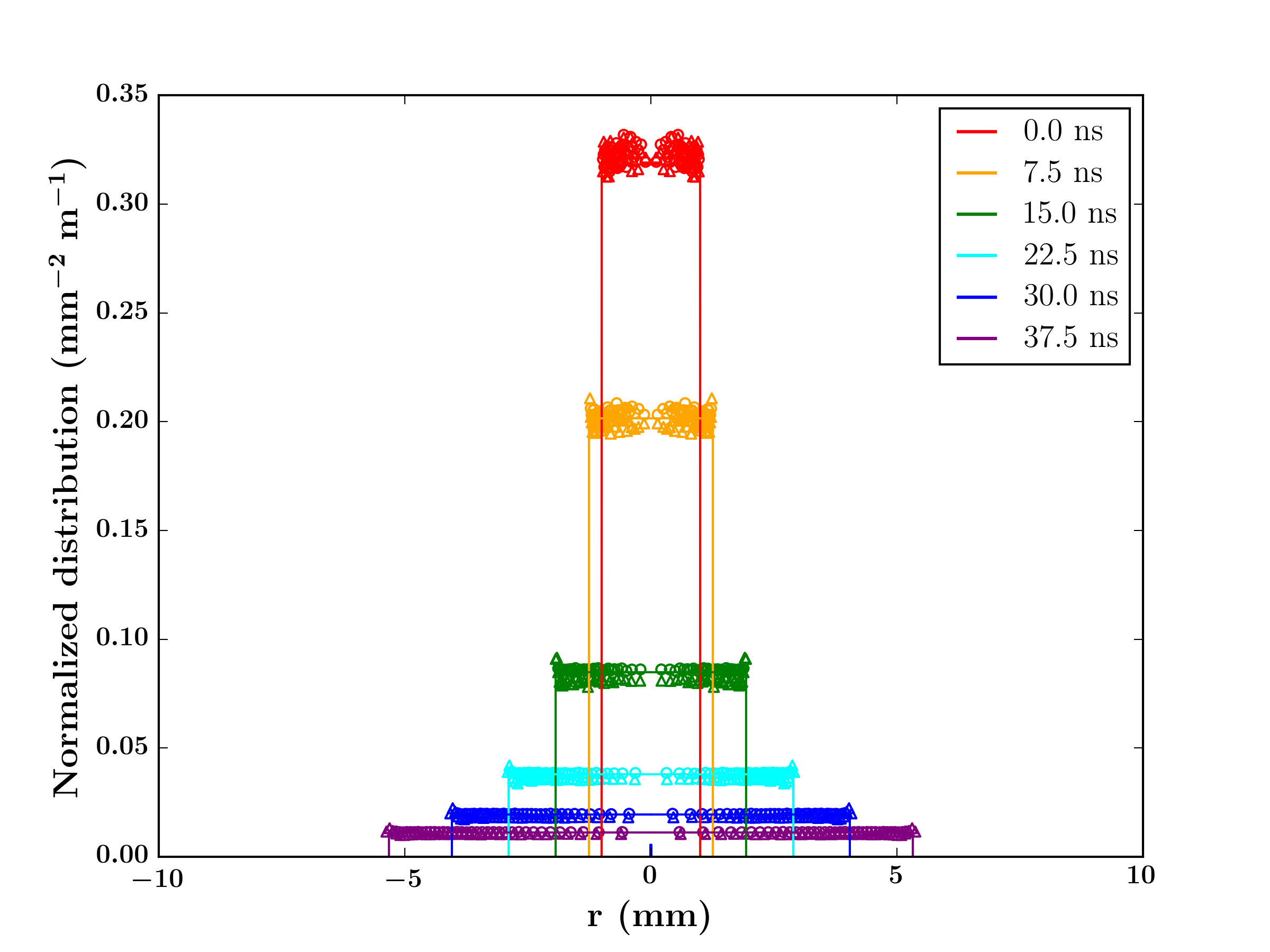}}&
    \subfloat[spherical symmetry uniform]{\includegraphics[width=0.45\textwidth]{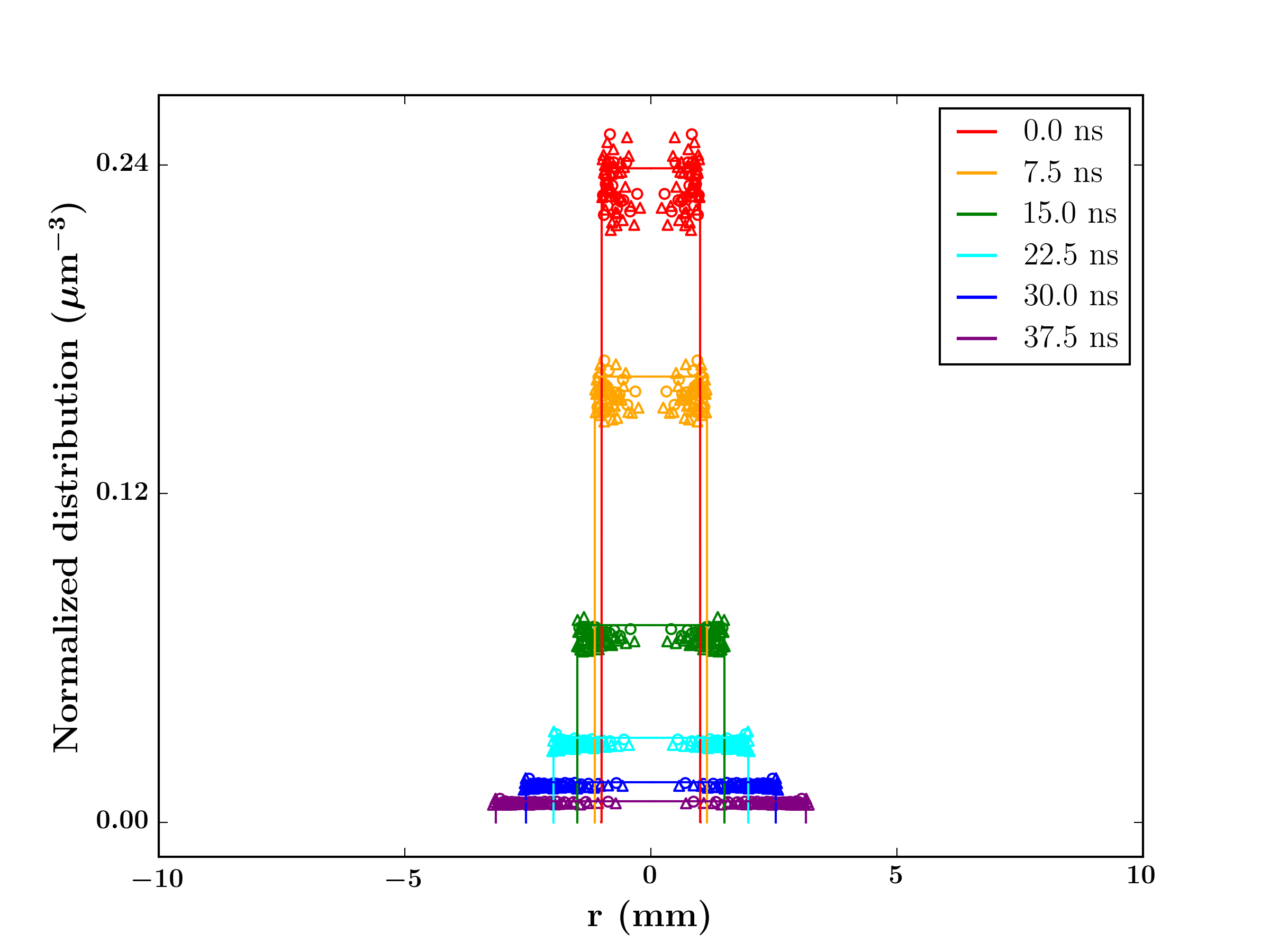}}
    \\
    \subfloat[cylindrical symmetry Gaussian]{\includegraphics[width=0.45\textwidth]{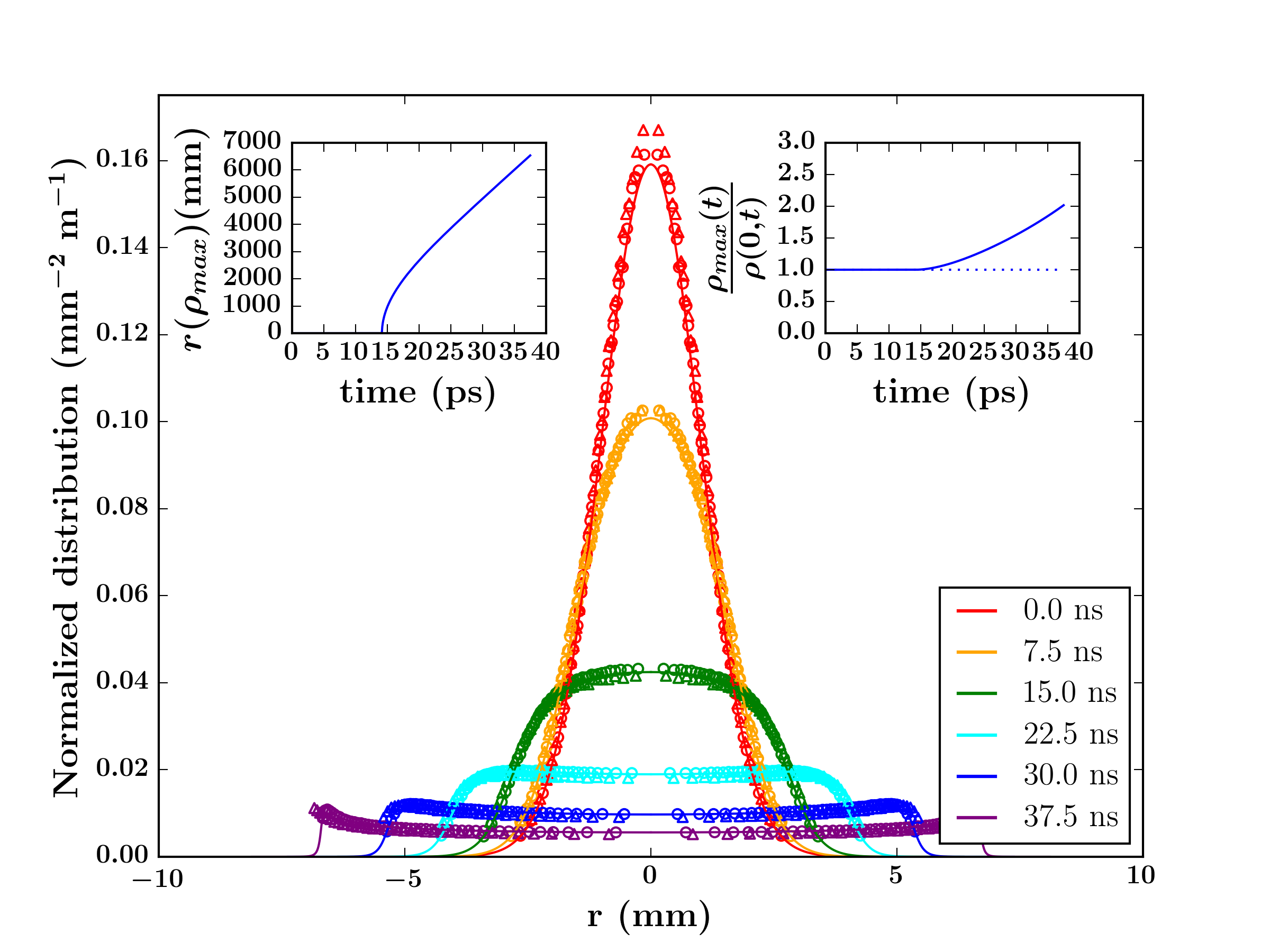}}&
    \subfloat[spherical symmetry Gaussian]{\includegraphics[width=0.45\textwidth]{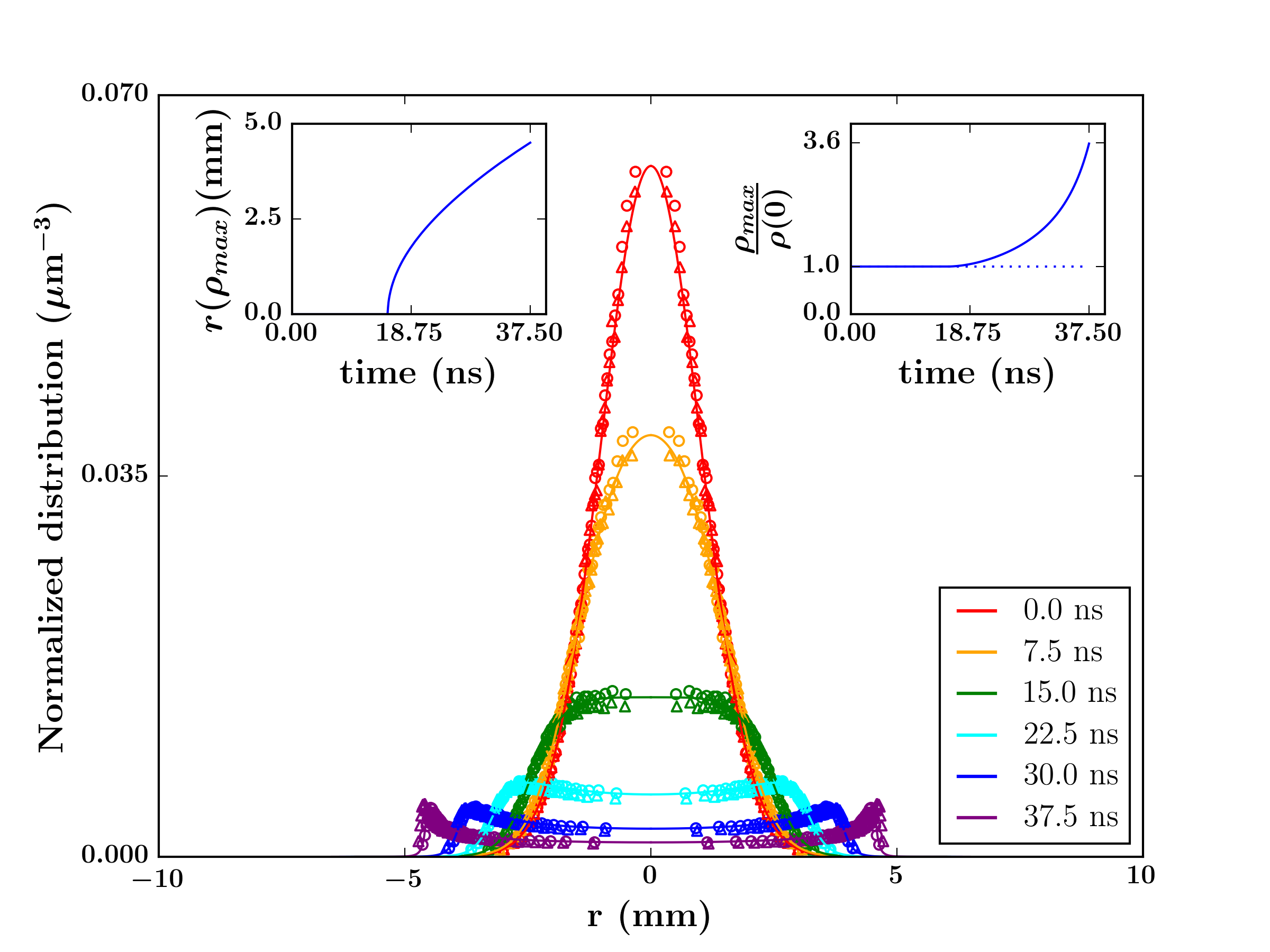}}
  \end{tabular}
\caption{\label{fig:density evolution} Analytical (solid line), PIC (circles), and N-particle (triangles) results of the normalized density 
evolution of (a.,c.) cylindrically and (b., d.) spherically symmetric 
(a., b.) uniform and (c.,d.) Gaussian distributions with $R = \sigma_r = 1 $ mm and  $N$ of $1.875 \times 10^7$, $2 \times 10^4$, $3.75 \times 10^7$,
and $10^5$ electrons, respectively.  The sub-graph
in the upper left corner of (c.,d.) shows 
the analytic position of max density as a function of time, and the sub-graph in the upper right of (c.d.) shows the analytic ratio of the
max density to the density at the minimum r value both.  The corresponding analytic ratio for the uniform distribution is shown in this sub-graph
as a dashed horizontal line at 1.   Unsurprisingly, the 
PIC results and the analytical results, both mean-field models, are in almost perfect agreement, and the N-particle results
are in surprisingly good agreement as well.
Notice that the models predict peak formation on a time-scale dependent on the initial plasma frequency similar to the peak formation seen
in the N-particle disc-like density evolution seen in Fig. (\ref{fig:distribution substructure}(e)) and detailed in Fig. (\ref{fig:emergence time}).}
\end{figure*} 

\section{Uniform and Gaussian Evolutions: Theory and Simulation}
In this section, the mean field predictions are compared to the N-particle and PIC simulations.
First we present the evolution of the initially-at-rest cylindrically- and spherically- symmetric uniform distribution of
$1.875 \times 10^7$ and $2 \times 10^4$ electrons within radii's of 1 mm (see Fig. (\ref{fig:density evolution}(a,b)).  Note, in this fairly trivial case, crossover should not 
occur and the analytic results should be valid mean-field-results for all time.  
Since $\rho_0 = {\bar{\rho}}_0$ in this case, $D_d(r_0) = 0$ and Eq. (\ref{eq:general density evolution})
reduces to
\begin{align}
  \rho(r;t) = \left(\frac{r_0}{r}\right)^d \rho_0(r_0)\label{eq:uniform evolution}
\end{align}
Notice that $r$ can be solved for a specific time using Eq. (\ref{eq:2D time}) or Eq. (\ref{eq:3D time}),
depending on whether we are examining the cylindrically- or spherically- symmetric case, respectively,
and due 
to ${\bar{\rho}}_0$'s independence from $r_0$, these equations need only be solved once 
for a give time to describe all $r$. Therefore, we may write $r = \alpha(t) r_0 \equiv \alpha r_0$, 
where $\alpha$ is independent of $r_0$, and
we immediately see that Eq. (\ref{eq:uniform evolution}) can be written as 
$\rho(r;t) = \alpha^d \rho_0(r_0)$ suggesting that
the density simply scales with time as generally recognized by the community. 
We solve for $\alpha$ at 6 times, and present a comparison with both PIC and 
N-particle cylindrically- symmetric and spherically-symmetric simulations in 
Fig. (\ref{fig:density evolution}(a,b)). As can be seen, despite the presence of initial density 
fluctuations arising from sampling, the simulated results follow the analytic results exceedingly well. 
Specifically, the distributions simply expand while remaining essentially uniform, and the 
analytic mean field formulation correctly calculates the rate of this expansion. 
While this comparison is arguably trivial, it is reassuring to see that our general equation 
reduces to a form that captures these dynamics.

Less trivial is the evolution of Gaussian distributions. We simulated 
$3.75 \times 10^7$ and $10^5$ electrons for the cylindrical and spherical cases, respectively,  using $\sigma_r =1 ~$ mm.  
Solving for the minimum crossover time, 
we get approximately 44 ns for each distribution.  Therefore, we simulate for 37.5 ns, which is well before any crossover events.  

For the Gaussian distributions we  
introduce the scaled radius variables $s = \frac{r}{\sqrt{2} \sigma_r}$ and 
$s_0 = \frac{r_0}{\sqrt{2}\sigma_r}$, so that from Eq. (25)  for the cylindrical and spherical cases we have,
\begin{align}
   D_2(s_0) &= \frac{(1 + s_0^2) e^{-s_0^2} - 1}{1 - e^{-s_0^2}}\\
  D_3(s_0) &= \frac{(2 s_0^3 + 3 s_0)e^{-s_0^2} - \frac{3 \sqrt{\pi}}{2} \text{erf}(s_0)}{\sqrt{\pi}\text{erf}(s_0) - 2 s_0 e^{-s_0^2}}
\end{align}

where erf is the well known error function. Putting these expressions into 
Eq. (\ref{eq:general density evolution}) we find for the cylindrical and 
\begin{widetext}
\noindent spherical cases respectively
\begin{align}
  \rho(s;t) &= \frac{\frac{s_0^2}{\pi s^2} e^{-s_0^2}}{1 + 2  \frac{(1 + s_0^2) e^{-s_0^2} - 1}{1 - e^{-s_0^2}} \sqrt{\ln\left(\frac{s}{s_0}\right)} F\left(\sqrt{\ln\left(\frac{s}{s_0}\right)}\right)}\\
  \rho(s;t) &= \frac{\frac{s_0^3}{\pi^{\frac{3}{2}} s^3} e^{-s_0^2}}{1 + \frac{(2 s_0^3 + 3 s_0)e^{-s_0^2} - \frac{3 \sqrt{\pi}}{2} \text{erf}(s_0)}{\sqrt{\pi}\text{erf}(s_0) - 2 s_0 e^{-s_0^2}} \left(\frac{s_0}{s}\sqrt{1 - \frac{s_0}{s}} \tanh^{-1}\left(\sqrt{1 - \frac{s_0}{s}}\right) + 1 - \frac{s_0}{s}\right)}
\end{align}

To find $r(t)/r_0$ we solve Eq. (\ref{eq:2D time}) or Eq. (\ref{eq:3D time}),
depending on whether we are examining the cylindrically- or 
spherically- symmetric cases, respectively; and 
for $\frac{r}{r_0}$, and for every time step, we calculate the
predicted distribution at 5000 positions, $r$, corresponding to 5000 initial positions, $r_0$, 
evolved to time $t$. As can be seen in 
Fig. (\ref{fig:density evolution}), both \end{widetext}
the cylindrically- and spherically- symmetric 
Gaussian distributions develop peaks similar to those seen in the simulations of expanding pancake bunches 
described in the first section of the paper.  
As can be seen in Fig. (\ref{fig:density evolution}), both the PIC and the N-particle 
results match the analytical results very well. Notice, the primary differences 
between the cylindrically- and spherically- symmetric evolutions is in their rate of width 
expansion and 
the sharpness of the peak that forms, and both of these facets are captured 
by the analytic models. 

\section{Conclusions}

In this work, we have shown that a shock occurs in  the transverse, but not longitudinal, 
direction during expansion of pancake-like charged particle distributions typical of those use 
in ultrafast electron microscope (UEM) systems. 

Fluid models for arbitrary initial 
distributions, Eq. (\ref{eq:1D density evolution subbed}), a generalization of a model 
already in the literature, showed that the formation of such a shock should 
not occur for any cold initial distribution in one dimension.  This result is consistent with the finding that 
typically no shock is visible in the longitudinal direction dynamics of UEM bunches; however 
by tuning the initial velocity distribution it should be possible to generate a dynamic shock. 

We generalized the fluid theory to cylindrical and 
spherical symmetries deriving implicit evolution equations for the charge density distributions
Eq. (\ref{eq:general density evolution}). We analyzed these models for the advent of 
particle crossover, which occur for some distributions even when the initial distribution is 
cold due to the behavior of the Coulomb force in higher dimensions; and we found that 
the time scales associated with the space charge expansion are proportional to the plasma period. 
One interesting detailed observation is that in the case of cylindrical 
symmetry, the pre-factor of $\frac{\tau_p}{\pi}$ of Eq. (\ref{eq:2D time}) is roughly 0.3 
while for the spherical symmetric case, 
corresponding prefactor in Eq. (\ref{eq:3D time}),  $\sqrt{\frac{3}{2}}\frac{\tau_p}{2\pi}$, is roughly
0.2 plasma periods. Interestingly  beam relaxation has been independently found 
to occur at roughly 0.25 the plasma period\cite{Wangler:1985_emittance_relaxation}, 
which falls directly in the middle of our cylindrically and spherically symmetric models. 
The analytic theory predicts that emergence of a shock is distribution 
dependent, and as expected, a uniform initial distribution does not produce a shock. 
However we showed that electron bunches that are initially Gaussian distributed
produce a shock well before the advent of particle crossover indicating that the emergence 
of a shock is well described by fluid models presented here.  
This  is consistent with the observation of a shock in N-particle simulations 
of the transverse expansion of UEM pancake bunches (see Figs. 1-4).

To our knowledge, we have presented the first analytic derivation of the cold, single-species, non-neutral 
density evolution equations for cylindrical and spherical symmetries. These equations are 
general enough to handle any distribution under these symmetries, and can be 
used across specialties from accelerator technology, to electronics, to astrophysics. While 
simulation methods, like the N-particle and PIC codes used here are general tools, 
the insights provided by these simple analytical equations should provide fast and easy 
first-approximations for a number of calculations; while providing physical insights and parameter 
dependences that are more difficult 
to extract from purely computational studies.  

The analysis presented here has been carried out for 
the non-relativistic regime; which is only valid for cases of sufficiently low density 
where the shock occurs prior to the electrons achieving relativistic velocities. For higher densities or  
other physical situations where the bunch becomes relativistic more quickly than the formation
of this shock, a relativistic analysis is needed. On the other hand,
for sufficiently high densities, i.e. approaching $10^7$ or more electrons in the pancake
geometry used in this manuscript and 
typical in the UE field, relativistic effects in the transverse direction become important
and need to be considered. The extension to fully relativistic cases will be addressed 
in future work.

We point out that Child-Langmuir current should not have these dynamic shocks except at the 
onset of the current before the steady-state condition sets in. Previous studies note the  ``hollowing'' 
of a steady-state beam due to fringe field effects\cite{Luginsland:1996_child_langmuir_2d}, 
but  a steady state Child-Langmuir 
current is largely independent of emission parameters; so that this hollowing effect is not 
dynamical,  but part of the continuous emission process itself, and is therefore a very different mechanism 
than the dynamic shocks we see here.  It would be interesting to study the combined 
effects of steady state beam hollowing and dynamic shock formation in pancake bunches 
to determine if the combination of these processes provides new opportunities for 
optimization of beam properties. 

The analytic models presented here treat free expansion whereas most applications have 
lattice elements to confine the bunches. Substantial work, in particular  the particle-core model,
 has been very successful at predicting 
transverse particle halos of 
beams\cite{Gluckstern:1994_analytic_halo,Wangler:1998_particle_core_review}. This 
model assumes a uniform-in-space beam-core density called a 
Kapchinsky-Vladimirsky (KV) distribution due to its ease of theoretical
treatment. Such an assumption is supported by the analysis presented here as we find that 
the distribution within the shock is nearly uniform. However, the particle-core models do 
not treat the initial distribution as having a large density on the periphery.  It would be interesting 
to revisit such treatments with this new perspective although we would like to point out that 
the main effect the particle-core model attempts to capture, halos, occur even after 
aperturing the beam\cite{Gluckstern:1994_analytic_halo}.   Specifically, it should be 
possible to examine the effect of radial-focussing fields on the evolution of the 
three-dimensional distributions we have investigated here.

The experimental work that motivated this analysis, \cite{Williams:2017_transverse_emittance}, 
not only predicted a shock but also a
correlated decrease in brightness near the periphery.  We emphasize that the mean-field equations
used here explain the density shock only and do not provide a quantitative theory of the emittance and
the Coulomb cooling achieved by removing the electrons in the shock.  
Specifically, the true emittance in the analytic models presented here remains zero for all time
as all particles at a radius $r$ have velocity  
$\sqrt{\frac{2}{3}}\frac{r_0}{{\bar{\omega}}_{p,0}} \sqrt{1 - \frac{r_0}{r}}$ resulting 
in zero local spread in velocity space.  This perfect relationship between velocity
and position means that the true emittance is zero even if the relation is non-linear;
however, in such a non-linear chirp case, the rms emittance will not remain zero despite the
true emittance being zero.   Moreover, the analytic model 
does capture some of the rms emittance growth as a change
 in the distribution has a corresponding change on the variance measures used to determine the rms emittance. 
Specifically, a Gaussian distribution should have especially large emittance growth 
due to its evolution 
to a bimodal distribution, a distribution that is 
specifically problematic for variance measures. Such a large change in the emittance of the 
transverse Gaussian  profile has been seen by 
Luiten\cite{Luiten:2004_uniform_ellipsoidal} experimentally and us 
computationally\cite{Portman:2013_computational_characterization, Portman:2014_image_charge}.  
On the other hand, 
the perfectly uniform 
distribution does not change its distribution throughout its evolution and therefore should have
zero rms emittance growth as the chirp exactly cancels out the expansion of the pulse
at all times.  Moreover, 
Luiten et al. found experimentally that the uniform distribution does have an increase in 
emittance although less than the Gaussian casel\cite{Luiten:2004_uniform_ellipsoidal}, 
an observation that is corroborated by our own work with PIC and N-particle 
calculations\cite{Portman:2013_computational_characterization, Portman:2014_image_charge}. 
 The analytic formulation of mean-field theory presented here provides new avenues to 
treating emittance growth, by treating fluctuations to these equations in a systematic manner. 
This analysis will be presented elsewhere. 

\textit{\textbf{Acknowledgment}}
This work was supported by NSF Grant 1625181, the College of Natural Science, 
the College of Communication Arts and Sciences, and the 
Provost's office at Michigan State University. Computational resources were provided by the 
High Performance Computer Center at MSU. We thank Martin Berz 
and Kyoko Makino for their help with employing COSY as an N-particle code.

\appendix
\section{1D Density Derivative}\label{ap:1D appendix}

In the main text, we argued
\begin{align}
  \rho(z)=  \rho_0(z_0) \left({dz\over dz_0}\right)^{-1}
\end{align}
To determine the slope of the density, we take the derivative with respect to the z coordinate
\begin{align}
  \frac{d}{dz} \rho(z) &= \frac{d}{dz_0} \left(\rho_0(z_0) \left(\frac{dz}{dz_0}\right)^{-1}\right)\left(\frac{dz}{dz_0}\right)^{-1}\nonumber\\
                                &= \frac{d}{dz_0} \left(\rho_0(z_0)\right) \left(\frac{dz}{dz_0}\right)^{-2} -  \rho_0(z_0) \left(\frac{dz}{dz_0}\right)^{-3} \frac{d^2z}{dz_0^2}\nonumber\\
                                &= \frac{\frac{d}{dz_0} \left(\rho_0(z_0)\right) \frac{dz}{dz_0} - \rho_0(z_0)\frac{d^2z}{dz_0^2}}{\left(\frac{dz}{dz_0}\right)^{3}}\label{eq:generic z derivative}
\end{align}
For the sake of conciseness, denote $\rho_0 = \rho_0(z_0)$, $\rho' = \frac{d}{dz} \rho(z)$, $\rho_0' = \frac{d}{dz_0} \rho_0$, $v_0' = \frac{d v_0}{d z_0}$, 
and $v_0'' = \frac{d^2 v_0}{d z_0^2}$.  
From the main text, we have
\begin{align}
  \frac{dz}{dz_0} &= 1 + v_0' t +  \frac{q}{2 m \epsilon_0} \rho_0(z_0)t^2
\end{align}
and from this it is straightforward to show 
\begin{align}
  \frac{d^2z}{dz_0^2} &= v_0'' t +  \frac{q}{2 m \epsilon_0} \rho_0't^2
\end{align}
Subbing this back into Eq. (\ref{eq:generic z derivative}), we get
\begin{align}
  \rho' &= \frac{\rho_0'\left(1 + v_0' t +  \frac{q}{2 m \epsilon_0} \rho_0t^2\right) - \rho_0\left(v_0'' t +  \frac{q}{2 m \epsilon_0} \rho_0't^2\right)}{\left(1 + v_0' t +  \frac{q}{2 m \epsilon_0} \rho_0t^2\right)^3}\nonumber\\
          &= \frac{ \rho_0'\left( 1  + v_0' t\right) - \rho_0v_0''t }{\left(1 + v_0' t +  \frac{q}{2 m \epsilon_0} \rho_0t^2\right)^3}\label{eq:1D rho derivative appendix} 
\end{align}

\section{Derivation of Time-location Relations}\label{ap:time-location appendix}
\subsection{Integral form}
Starting with the relativistic expression for change in particle energy derived in the main text
\begin{align}
  \text{cyl:  }E(t) - E(0) &= \frac{q Q_{tot} \lambda_0}{2 \pi \epsilon_0} ln\left(\frac{r}{r_0}\right)\label{eq:2D energy}\\
  \text{sph: }E(t) - E(0) &= \frac{q Q_{tot} P_0}{4 \pi \epsilon_0} \left(\frac{1}{r_0} - \frac{1}{r}\right)\label{eq:3D energy}
\end{align}
we approximate the energy change with a change in non-relativistic kinetic energy starting from rest
\begin{align}
  \text{cyl: }\frac{1}{2} m v^2 &= \frac{q Q_{tot} \lambda_0}{2 \pi \epsilon_0} ln\left(\frac{r}{r_0}\right)\label{eq:2D kinetic energy}\\
   \text{sph: }\frac{1}{2} m v^2 &= \frac{q Q_{tot} P_0}{4 \pi \epsilon_0} \left(\frac{1}{r_0} - \frac{1}{r}\right)\label{eq:3D kinetic energy}
\end{align}
where $v = \frac{d r}{dt}$ are the velocity of the particle at time $t$ in the two or one of 
the three dimensional
models, respectively, with the appropriate definition of $r$.  Solving these equations 
for the velocity at time $t$, we get
\begin{align}
  \text{cyl: } \frac{d r}{d t} &= \sqrt{\frac{q Q_{tot} \lambda_0}{\pi m \epsilon_0} ln\left(\frac{r}{r_0}\right)}\label{eq:2D velocity}\\
  \text{sph: } \frac{d r}{d t} &= \sqrt{\frac{q Q_{tot} P_0}{2 \pi m \epsilon_0} \left(\frac{1}{r_0} - \frac{1}{r }\right)}\label{eq:3D velocity}
\end{align}  
Separating the variables and integrating, we obtain
\begin{align}
  \text{cyl: } t &= \int_{r_0}^{r} \frac{d\tilde{r}}{ \sqrt{\frac{q Q_{tot} \lambda_0}{\pi m \epsilon_0} ln\left(\frac{\tilde{r}}{r_0}\right)}}\label{eq:2D time integral}\\
  \text{sph: } t &= \int_{r_0}^{r} \frac{d\tilde{r}}{\sqrt{\frac{q Q_{tot} P_0}{2 \pi m \epsilon_0} \left(\frac{1}{r_0} - \frac{1}{\tilde{r}}\right)}}\label{eq:3D time integral}
\end{align}
Defining
$a = \frac{q Q_\text{tot}}{\pi m \epsilon_0}$, we rewrite 
Eq. (\ref{eq:2D time integral}) and Eq. (\ref{eq:3D time integral}) as
\begin{align}
  \text{cyl: }t &= \int_{r_0}^{r} \frac{d\tilde{r}}{ \sqrt{a \lambda_0 ln\left(\frac{\tilde{r}}{r_0}\right)}}\label{eq:2D time integral subbed}\\
  \text{sph: }t &= \int_{r_0}^{r} \frac{d\tilde{r}}{\sqrt{\frac{aP_0}{2r_0}}\sqrt{1 - \frac{r_0}{\tilde{r} }}}\label{eq:3D time integral subbed}
\end{align} 

\subsection{Cylindrically-symmetric integral solution}
We solve the sylindrically-symmetric integral first.  Define $\tilde{u} = \sqrt{a \lambda_0 ln\left(\frac{\tilde{r}}{r_0}\right)}$.  Solving this equation for $\tilde{r}$ in terms of
$\tilde{u}$, we see that $\tilde{r} = r_0 e^{\frac{\tilde{u}^2}{a \lambda_0}}$.  It is
also straightforward to see that 
\begin{align}
  d \tilde{u} &= \frac{1}{2} \frac{1}{ \sqrt{a \lambda_0 ln\left(\frac{\tilde{r}}{r_0}\right)}} \frac{a \lambda_0}{\tilde{r}} d \tilde{r}\nonumber\\
                  &= \frac{1}{ \sqrt{a \lambda_0 ln\left(\frac{\tilde{r}}{r_0}\right)}} \frac{a \lambda_0}{2 r_0} e^{\frac{-\tilde{u}^2}{a \lambda_0}} d \tilde{r} \nonumber
\end{align}
Applying this change of coordinates to Eq. (\ref{eq:2D time integral subbed}), we get
\begin{align}
  \text{cyl: } t &= \int_{0}^{u} \frac{2 r_0}{a \lambda_0}e^{\frac{\tilde{u}^2}{a \lambda_0}} d \tilde{u}\nonumber\\
                    &= \frac{2 r_0}{a \lambda_0}e \int_{0}^{u} e^{\frac{\tilde{u}^2}{a \lambda_0}} d \tilde{u}\nonumber\\
                    &= \frac{2 r_0}{\sqrt{a \lambda_0}} \int_{0}^{w} e^{\tilde{w}^2} d \tilde{w}\label{eq:2D time integral in w}
\end{align}
where $u = \sqrt{a \lambda_0 ln\left(\frac{r}{r_0}\right)}$, 
$\tilde{w} = \frac{\tilde{u}}{\sqrt{a \lambda_0}}$, and $w = \sqrt{ln\left(\frac{r}{r_0}\right)}$.  The remaining integral, 
$\int_{0}^{w} e^{\tilde{w}^2} d \tilde{w}$ can be written in terms of the well-studied Dawson function, $F(\cdot)$:
\begin{align}
\int_{0}^{w} e^{\tilde{w}^2} d \tilde{w} &= e^{w^2}F(w) \nonumber\\
                                                           &= \frac{r}{r_0} F\left(\sqrt{ln\left(\frac{r}{r_0}\right)}\right)\label{eq:integral as Dawson}
\end{align}
Subbing Eq. (\ref{eq:integral as Dawson}) back into Eq. (\ref{eq:2D time integral in w}) gives us our time-position relation
\begin{align}\label{eq:2D time vs r}
  \text{cyl: } t &= \frac{2 r}{\sqrt{a \lambda_0}} F\left(\sqrt{ln\left(\frac{r}{r_0}\right)} \right)\nonumber\\
                     &= 2 \frac{r}{r_0} \sqrt{\frac{\pi r_0^2 m \epsilon_0}{q Q_\text{tot}\lambda_0}} F\left(\sqrt{ln\left(\frac{r}{r_0}\right)} \right)\nonumber\\
                     &= 2  \frac{r}{r_0} \sqrt{\frac{m \epsilon_0}{q Q_\text{tot}\frac{\lambda_0}{\pi r_0^2}}}F\left(\sqrt{ln\left(\frac{r}{r_0}\right)} \right)\nonumber\\
                     &= 2  \frac{r}{r_0} \sqrt{\frac{m \epsilon_0}{q Q_\text{tot}{\bar{\rho}}_0}}F\left(\sqrt{ln\left(\frac{r}{r_0}\right)} \right)\nonumber\\
                     &= \frac{2}{{\bar{\omega}}_{p,0}}  \frac{r}{r_0} F\left(\sqrt{ln\left(\frac{r}{r_0}\right)} \right)\nonumber\\
                     &= \frac{{\bar{\tau}}_{p,0}}{\pi} \frac{r}{r_0} F\left(\sqrt{ln\left(\frac{r}{r_0}\right)} \right)
\end{align}
where ${\bar{\rho}}_0 = \frac{\lambda_0}{\pi r_0^2}$ and ${\bar{\omega}}_{p,0} = \sqrt{\frac{q Q_\text{tot}{\bar{\rho}}_0}{m \epsilon_0}} = \frac{2 \pi}{{\bar{\tau}}_{p,0}}$.

\subsection{Spherically-symmetric Integral Solution}
We solve the spherically-symmetric integral with an analogous approach.  Define $\tilde{u} = \sqrt{1 - \frac{r_0}{\tilde{r} }}$ and solving for $\tilde{r}$ gives
$\tilde{r} = \frac{r_0}{1-\tilde{u}^2}$.  Thus
\begin{align}
  d \tilde{u} &= \frac{1}{2} \frac{1}{\sqrt{1 - \frac{r_0}{\tilde{r} }}}\frac{r_0}{\tilde{r}^2 } d \tilde{r}\nonumber\\
                  &= \frac{1}{\sqrt{1 - \frac{r_0}{\tilde{r} }}} \frac{(1 - \tilde{u}^2)^2}{2 r_0} d\tilde{r}\nonumber
\end{align}
Applying this change of coordinates to Eq. (\ref{eq:3D time integral subbed}) with
$u = \sqrt{1 - \frac{r_0}{r}}$, we get
\begin{align}
  \text{sph: }t &= \sqrt{\frac{2 r_0}{aP_0}}\int_{0}^{u} \frac{2 r_0}{(1 - \tilde{u}^2)^2} d \tilde{u}\nonumber\\
                    &=  2 \sqrt{\frac{2 r_0^3}{aP_0}} \int_{0}^{u} \frac{1}{(1 - \tilde{u}^2)^2} d \tilde{u} \nonumber\\   
                    &=  2 \sqrt{\frac{2 r_0^3}{aP_0}} \left( \frac{1}{2} \tanh^{-1} \left(\tilde{u}\right) + \frac{1}{2 } \frac{\tilde{u}}{1 - \tilde{u}^2}\right) \bigg \rvert_{\tilde{u} = 0}^{\tilde{u} = \sqrt{1 - \frac{r_0}{r}}}\nonumber\\
                    &= \sqrt{\frac{2 r_0^3}{a P_0}} \left( \tanh^{-1} \left(\sqrt{1 - \frac{r_0}{r}}\right) + \frac{\sqrt{1 - \frac{r_0}{r}}}{1 - 1 + \frac{r_0}{r}}\right)\nonumber\\
                    &= \sqrt{\frac{2 \pi r_0^3 m \epsilon_0}{q Q_\text{tot} P_0}} \left( \tanh^{-1} \left(\sqrt{1 - \frac{r_0}{r}}\right) + \frac{r}{r_0}\sqrt{1 - \frac{r_0}{r}}\right)\nonumber\\
                    &= \sqrt{\frac{2}{3}} \sqrt{\frac{m \epsilon_0}{q Q_\text{tot} \frac{P_0}{\frac{4}{3}\pi r_0^3}}} \left( \tanh^{-1} \left(\sqrt{1 - \frac{r_0}{r}}\right)\right.\nonumber\\
                    &\quad\quad \left.+ \frac{r}{r_0}\sqrt{1 - \frac{r_0}{r}}\right)\nonumber\\
                    &= \sqrt{\frac{2}{3}} \sqrt{\frac{m \epsilon_0}{q Q_\text{tot} {\bar{\rho}}_0}} \left( \tanh^{-1} \left(\sqrt{1 - \frac{r_0}{r}}\right)\right.\nonumber\\
                    &\quad\quad \left. + \frac{r}{r_0}\sqrt{1 - \frac{r_0}{r}}\right)\nonumber\\
                    &= \sqrt{\frac{2}{3}} \frac{1}{{\bar{\omega}}_{p,0}} \left( \tanh^{-1} \left(\sqrt{1 - \frac{r_0}{r}}\right) + \frac{r}{r_0}\sqrt{1 - \frac{r_0}{r}}\right)\nonumber\\
                    &= \sqrt{\frac{2}{3}} \frac{{\bar{\tau}}_{p,0}}{2 \pi} \left( \tanh^{-1} \left(\sqrt{1 - \frac{r_0}{r}}\right) + \frac{r}{r_0}\sqrt{1 - \frac{r_0}{r}}\right)\label{eq:3D time vs r}
\end{align}
where the solution to the integral was obtained with Mathematica's online tool\cite{Mathematica} and
where ${\bar{\rho}}_0 = \frac{P_0}{\frac{4}{3}\pi r_0^3}$ and ${\bar{\omega}}_{p,0} = \frac{q Q_\text{tot}{\bar{\rho}}_0}{m \epsilon_0} = \frac{2 \pi}{{\bar{\tau}}_{p,0}}$.

\section{Derivation of Derivatives with Respect to Initial Position}\label{ap:spatial derivatives}
As noted in the main text, much of the physics of distribution evolution in our models is captured in the term $\frac{dr}{dr_0}$.  
The procedure to derive the expressions for this derivative is to take the derivative of Eq. (\ref{eq:2D time vs r}) and Eq.
(\ref{eq:3D time vs r}).  We do this mathematics here.

\subsection{The Cylindrically-symmetric Derivative}
We begin by re-writing $t$ from Eq. (\ref{eq:2D time vs r}) as
\begin{align}
  t &= 2  r \sqrt{\frac{1}{a \lambda_0}} F\left(y \right)
\end{align}
where $y = \sqrt{ln\left(\frac{r}{r_0}\right)} = \sqrt{\ln \left(r \right) - \ln\left({r_0}\right)}$.  So
\begin{align}
  \frac{dy}{dr_0} &= \frac{1}{2y} \left(\frac{1}{r} \frac{d r}{d r_0} - \frac{1}{r_0} \right)\label{eq:derivative 2D sqrt term} 
 \end{align}
The Dawson function has the property $\frac{d}{dy} F(y) = 1 - 2 y F(y) = \left(\frac{1}{F(y)} - 2 y\right) F(y)$, and with the chain rule this becomes
 $\frac{d}{dr_0} F(y) = \left(\frac{1}{F(y)} - 2 y\right)  \frac{dy}{dr_0} F(y)$.  Using Eq. (\ref{eq:derivative 2D sqrt term}), this becomes 
\begin{align} 
  \frac{d}{dr_0} F\left(y\right) &= \left(\frac{1}{F(y)} - 2 y\right)\frac{F(y)}{2y} \left(\frac{1}{r} \frac{d r}{d r_0} - \frac{1}{r_0} \right)\nonumber\\
                                              &=F(y) \left(\frac{1}{2 y F(y)} - 1\right)\left(\frac{1}{r} \frac{d r}{d r_0} - \frac{1}{r_0} \right)\label{eq:derivative our Dawson} 
\end{align}
Also, note
\begin{align}
  \frac{d}{dr_0}\frac{1}{\sqrt{\lambda_0}} &= -\frac{1}{2}\frac{1}{(\lambda_0)^{3/2}} \frac{d\lambda_0}{d r_0}\nonumber\\
                                                                   &= -\frac{1}{2 \sqrt{\lambda_0}} \frac{d \ln(\lambda_0)}{d r_0}\label{eq:derivative sqrt a lambda_0}
 \end{align}
So
\begin{align}
  0 &= \frac{dt}{dr_0}\nonumber\\
     &= \frac{t}{r} \frac{dr}{dr_0} - \frac{1}{2} t  \frac{d \ln(\lambda_0)}{d r_0} + t \left(\frac{1}{2 y F(y)} - 1\right)\left(\frac{1}{r} \frac{d r}{d r_0} - \frac{1}{r_0} \right)\nonumber\\
     &= \frac{t}{r} \frac{1}{2 y F(y)}\frac{dr}{dr_0} - t \left(\frac{1}{2} \frac{d \ln(\lambda_0)}{d r_0} + \left(\frac{1}{2 y F(y)} - 1\right) \frac{1}{r_0}\right)
\end{align}
which gives 
\begin{align}
  \frac{dr}{dr_0} &= 2 y F(y) r \left(\frac{1}{2} \frac{d \ln(\lambda_0)}{d r_0} + \left(\frac{1}{2 y F(y)} - 1\right) \frac{1}{r_0}\right)\nonumber\\
                         &= \frac{r}{r_0}\left( 1 + \left(\frac{r_0}{2} \frac{d \ln(\lambda_0)}{d r_0} - 1 \right) 2 y F(y) \right)\nonumber\\
                         &= \frac{r}{r_0}\left( 1 + \left(\frac{r_0}{2} \frac{2 \pi r_0 \rho_0}{\lambda_0}  - 1 \right) 2 y F(y) \right)\nonumber\\
                         &= \frac{r}{r_0}\left( 1 + \left(\frac{\rho_0}{\frac{\lambda_0}{\pi r_0^2}}  - 1 \right) 2 y F(y) \right)\nonumber\\ 
                         &= \frac{r}{r_0} \left(1 + \left(\frac{\rho_0}{\bar{\rho}_0} - 1\right) 2  \sqrt{\ln\left(\frac{r}{r_0}\right)}F\left(\sqrt{\ln\left(\frac{r}{r_0}\right)}\right)\right) 
\end{align}

\subsection{The Spherically-symmetric Derivatives}
We begin by re-writing $t$ from Eq. (\ref{eq:3D time vs r}) as
\begin{align}
  t &= \sqrt{\frac{2 r_0^3}{a P_0}} \left(\tanh^{-1} y + \frac{r}{r_0} y \right)
\end{align}
where $y = \sqrt{1 - \frac{r_0}{r}}$.  So
\begin{align}
  \frac{dy}{dr_0} &= \frac{1}{2y}\left(\frac{r_0}{r^2} \frac{dr}{dr_0} - \frac{1}{r}\right)\nonumber\\
                          &= - \frac{1}{2yr}\left(1- \frac{r_0}{r} \frac{dr}{dr_0}\right) 
\end{align}
Hence
\begin{align}
  \frac{d\tanh^{-1} y}{dr_0} &= \frac{1}{1 - y^2} \frac{dy}{dr_0}\nonumber\\
                                        &= \frac{r}{r_0} \left(- \frac{1}{2yr}\left(1- \frac{r_0}{r} \frac{dr}{dr_0}\right)\right)\nonumber\\
                                        &= - \frac{1}{2yr_0}\left(1- \frac{r_0}{r} \frac{dr}{dr_0}\right)
\end{align}
and 
\begin{align}
  \frac{d\left(\frac{r}{r_0} y \right)}{dr_0} &= \frac{y}{r_0} \frac{dr}{dr_0} - \frac{r}{r_0^2}y + \frac{r}{r_0} \frac{dy}{dr_0}\nonumber\\
                                                              &= \frac{1}{r_0}\left( \left(\frac{dr}{dr_0} - \frac{r}{r_0}\right)y -\frac{1}{2y}\left(1- \frac{r_0}{r} \frac{dr}{dr_0}\right)   \right)\nonumber\\
                                                              &= \frac{1}{2 y r_0}\left( 2 \left(\frac{dr}{dr_0} - \frac{r}{r_0}\right)\left(1 - \frac{r_0}{r}\right) - 1+ \frac{r_0}{r} \frac{dr}{dr_0}\right)\nonumber\\ 
\end{align}
Therefore
\begin{align}
  \frac{d\left(\tanh^{-1} y + \frac{r}{r_0} y \right)}{dr_0} &= \frac{1}{y r_0}\left( \left(\frac{dr}{dr_0} - \frac{r}{r_0}\right)\left(1 - \frac{r_0}{r}\right)\right.\nonumber\\
                            &\quad\quad \left. - 1+ \frac{r_0}{r} \frac{dr}{dr_0}\right)\nonumber\\ 
                                                                                    &= \frac{1}{y r_0}\left( \frac{dr}{dr_0} - \frac{r}{r_0}\right)\nonumber\\  
\end{align}
Also, similar to Eq. (\ref{eq:derivative sqrt a lambda_0}), $\frac{d}{dr_0}\frac{1}{\sqrt{P_0}} = -\frac{1}{2 \sqrt{P_0}} \frac{d \ln(P_0)}{d r_0}$.  Putting this together we have
\begin{align}
  0 &= \frac{dt}{dr_0}\nonumber\\
     &= \frac{3}{2} \frac{t}{r_0} - \frac{t}{2} \frac{d \ln(P_0)}{d r_0} +  \sqrt{\frac{2 r_0^3}{a P_0}} \frac{d\left(\tanh^{-1} y + \frac{r}{r_0} y \right)}{dr_0}\nonumber\\
     &= \frac{3}{2} \frac{t}{r_0} - \frac{t}{2} \frac{d \ln(P_0)}{d r_0} +  \sqrt{\frac{2 r_0^3}{a P_0}} \frac{1}{y r_0}\left( \frac{dr}{dr_0} - \frac{r}{r_0} \right)
\end{align}
Solving for $\frac{dr}{dr_0}$ we get
\begin{align}
  \frac{dr}{dr_0} &= \frac{r}{r_0} - \frac{3y}{2} \left(\tanh^{-1} y + \frac{r}{r_0} y \right)  \nonumber\\
                        &\quad\quad + \frac{y r_0}{2}\left(\tanh^{-1} y + \frac{r}{r_0} y \right)\frac{d \ln(P_0)}{d r_0}\nonumber\\
                         &= \frac{r}{r_0}  + \frac{3y}{2}\left( \frac{r_0}{3}\frac{d \ln(P_0)}{d r_0} - 1\right) \left(\tanh^{-1} y + \frac{r}{r_0} y \right)\nonumber\\
                         &=  \frac{r}{r_0} \left(1 + \frac{3}{2}\left(\frac{r_0}{3}\frac{4 \pi r_0^2 \rho_0}{P_0} - 1\right) \left(\frac{r_0}{r} y \tanh^{-1} y + y^2 \right)\right)\nonumber\\
                         &=  \frac{r}{r_0} \left(1 + \frac{3}{2}\left(\frac{\rho_0}{\frac{P_0}{\frac{4}{3}\pi r_0^3}} - 1\right)\left(\frac{r_0}{r} \sqrt{1 - \frac{r_0}{r}}  \tanh^{-1}\left(\sqrt{1 - \frac{r_0}{r}}\right)\right.\right.\nonumber\\
                         &\quad\quad \left.\left.  + 1 - \frac{r_0}{r} \right)\right)\nonumber\\
                         &= \frac{r}{r_0} \left(1 + \frac{3}{2}\left(\frac{\rho_0}{{\bar{\rho}}_0} - 1\right)\left(\frac{r_0}{r} \sqrt{1 - \frac{r_0}{r}}  \tanh^{-1}\left(\sqrt{1 - \frac{r_0}{r}}\right) \right.\right. \nonumber\\
                         &\quad\quad \left.\left. + 1 - \frac{r_0}{r} \right)\right)
\end{align}
\bibliographystyle{apsrev4-1}
\bibliography{CoulombDynamics}

\begin{thebibliography}{67}%
\makeatletter
\providecommand \@ifxundefined [1]{%
 \@ifx{#1\undefined}
}%
\providecommand \@ifnum [1]{%
 \ifnum #1\expandafter \@firstoftwo
 \else \expandafter \@secondoftwo
 \fi
}%
\providecommand \@ifx [1]{%
 \ifx #1\expandafter \@firstoftwo
 \else \expandafter \@secondoftwo
 \fi
}%
\providecommand \natexlab [1]{#1}%
\providecommand \enquote  [1]{``#1''}%
\providecommand \bibnamefont  [1]{#1}%
\providecommand \bibfnamefont [1]{#1}%
\providecommand \citenamefont [1]{#1}%
\providecommand \href@noop [0]{\@secondoftwo}%
\providecommand \href [0]{\begingroup \@sanitize@url \@href}%
\providecommand \@href[1]{\@@startlink{#1}\@@href}%
\providecommand \@@href[1]{\endgroup#1\@@endlink}%
\providecommand \@sanitize@url [0]{\catcode `\\12\catcode `\$12\catcode
  `\&12\catcode `\#12\catcode `\^12\catcode `\_12\catcode `\%12\relax}%
\providecommand \@@startlink[1]{}%
\providecommand \@@endlink[0]{}%
\providecommand \url  [0]{\begingroup\@sanitize@url \@url }%
\providecommand \@url [1]{\endgroup\@href {#1}{\urlprefix }}%
\providecommand \urlprefix  [0]{URL }%
\providecommand \Eprint [0]{\href }%
\providecommand \doibase [0]{http://dx.doi.org/}%
\providecommand \selectlanguage [0]{\@gobble}%
\providecommand \bibinfo  [0]{\@secondoftwo}%
\providecommand \bibfield  [0]{\@secondoftwo}%
\providecommand \translation [1]{[#1]}%
\providecommand \BibitemOpen [0]{}%
\providecommand \bibitemStop [0]{}%
\providecommand \bibitemNoStop [0]{.\EOS\space}%
\providecommand \EOS [0]{\spacefactor3000\relax}%
\providecommand \BibitemShut  [1]{\csname bibitem#1\endcsname}%
\let\auto@bib@innerbib\@empty
\bibitem [{\citenamefont {Arba{\~n}il}\ \emph {et~al.}(2014)\citenamefont
  {Arba{\~n}il}, \citenamefont {Lemos},\ and\ \citenamefont
  {Zanchin}}]{Arbanil:2014_charged_star_review}%
  \BibitemOpen
  \bibfield  {author} {\bibinfo {author} {\bibfnamefont {J.~D.}\ \bibnamefont
  {Arba{\~n}il}}, \bibinfo {author} {\bibfnamefont {J.~P.}\ \bibnamefont
  {Lemos}}, \ and\ \bibinfo {author} {\bibfnamefont {V.~T.}\ \bibnamefont
  {Zanchin}},\ }\href@noop {} {\bibfield  {journal} {\bibinfo  {journal}
  {Physical Review D}\ }\textbf {\bibinfo {volume} {89}},\ \bibinfo {pages}
  {104054} (\bibinfo {year} {2014})}\BibitemShut {NoStop}%
\bibitem [{\citenamefont {Maurya}\ \emph {et~al.}(2015)\citenamefont {Maurya},
  \citenamefont {Gupta}, \citenamefont {Ray},\ and\ \citenamefont
  {Chowdhury}}]{Maurya:2015_charged_sphere}%
  \BibitemOpen
  \bibfield  {author} {\bibinfo {author} {\bibfnamefont {S.}~\bibnamefont
  {Maurya}}, \bibinfo {author} {\bibfnamefont {Y.}~\bibnamefont {Gupta}},
  \bibinfo {author} {\bibfnamefont {S.}~\bibnamefont {Ray}}, \ and\ \bibinfo
  {author} {\bibfnamefont {S.~R.}\ \bibnamefont {Chowdhury}},\ }\href@noop {}
  {\bibfield  {journal} {\bibinfo  {journal} {The European Physical Journal C}\
  }\textbf {\bibinfo {volume} {75}},\ \bibinfo {pages} {389} (\bibinfo {year}
  {2015})}\BibitemShut {NoStop}%
\bibitem [{\citenamefont {Yousefi}\ \emph {et~al.}(2014)\citenamefont
  {Yousefi}, \citenamefont {Davis}, \citenamefont {Carmona-Reyes},
  \citenamefont {Matthews},\ and\ \citenamefont
  {Hyde}}]{Yousefi:2014_dust_aggregates}%
  \BibitemOpen
  \bibfield  {author} {\bibinfo {author} {\bibfnamefont {R.}~\bibnamefont
  {Yousefi}}, \bibinfo {author} {\bibfnamefont {A.~B.}\ \bibnamefont {Davis}},
  \bibinfo {author} {\bibfnamefont {J.}~\bibnamefont {Carmona-Reyes}}, \bibinfo
  {author} {\bibfnamefont {L.~S.}\ \bibnamefont {Matthews}}, \ and\ \bibinfo
  {author} {\bibfnamefont {T.~W.}\ \bibnamefont {Hyde}},\ }\href@noop {}
  {\bibfield  {journal} {\bibinfo  {journal} {Physical Review E}\ }\textbf
  {\bibinfo {volume} {90}},\ \bibinfo {pages} {033101} (\bibinfo {year}
  {2014})}\BibitemShut {NoStop}%
\bibitem [{\citenamefont {Bacci}\ \emph {et~al.}(2014)\citenamefont {Bacci},
  \citenamefont {Rossi} \emph {et~al.}}]{Bacci:2014_plasma_acceleration}%
  \BibitemOpen
  \bibfield  {author} {\bibinfo {author} {\bibfnamefont {A.}~\bibnamefont
  {Bacci}}, \bibinfo {author} {\bibfnamefont {A.}~\bibnamefont {Rossi}},  \emph
  {et~al.},\ }\href@noop {} {\bibfield  {journal} {\bibinfo  {journal} {Nuclear
  Instruments and Methods in Physics Research Section A: Accelerators,
  Spectrometers, Detectors and Associated Equipment}\ }\textbf {\bibinfo
  {volume} {740}},\ \bibinfo {pages} {42} (\bibinfo {year} {2014})}\BibitemShut
  {NoStop}%
\bibitem [{\citenamefont {Boine-Frankenheim}\ \emph {et~al.}(2015)\citenamefont
  {Boine-Frankenheim}, \citenamefont {Hofmann}, \citenamefont {Struckmeier},\
  and\ \citenamefont {Appel}}]{Boine:2015_intense_beams}%
  \BibitemOpen
  \bibfield  {author} {\bibinfo {author} {\bibfnamefont {O.}~\bibnamefont
  {Boine-Frankenheim}}, \bibinfo {author} {\bibfnamefont {I.}~\bibnamefont
  {Hofmann}}, \bibinfo {author} {\bibfnamefont {J.}~\bibnamefont
  {Struckmeier}}, \ and\ \bibinfo {author} {\bibfnamefont {S.}~\bibnamefont
  {Appel}},\ }\href@noop {} {\bibfield  {journal} {\bibinfo  {journal} {Nuclear
  Instruments and Methods in Physics Research Section A: Accelerators,
  Spectrometers, Detectors and Associated Equipment}\ }\textbf {\bibinfo
  {volume} {770}},\ \bibinfo {pages} {164} (\bibinfo {year}
  {2015})}\BibitemShut {NoStop}%
\bibitem [{\citenamefont {Whelan}\ \emph {et~al.}(2016)\citenamefont {Whelan},
  \citenamefont {Gierman}, \citenamefont {Holloway}, \citenamefont {Schmerge},
  \citenamefont {Keall},\ and\ \citenamefont {Fahrig}}]{Whelan:2016_MRI}%
  \BibitemOpen
  \bibfield  {author} {\bibinfo {author} {\bibfnamefont {B.}~\bibnamefont
  {Whelan}}, \bibinfo {author} {\bibfnamefont {S.}~\bibnamefont {Gierman}},
  \bibinfo {author} {\bibfnamefont {L.}~\bibnamefont {Holloway}}, \bibinfo
  {author} {\bibfnamefont {J.}~\bibnamefont {Schmerge}}, \bibinfo {author}
  {\bibfnamefont {P.}~\bibnamefont {Keall}}, \ and\ \bibinfo {author}
  {\bibfnamefont {R.}~\bibnamefont {Fahrig}},\ }\href@noop {} {\bibfield
  {journal} {\bibinfo  {journal} {Medical physics}\ }\textbf {\bibinfo {volume}
  {43}},\ \bibinfo {pages} {1285} (\bibinfo {year} {2016})}\BibitemShut
  {NoStop}%
\bibitem [{\citenamefont {Bernal}\ \emph {et~al.}(2016)\citenamefont {Bernal},
  \citenamefont {Beaudoin}, \citenamefont {Haber}, \citenamefont {Koeth},
  \citenamefont {Mo}, \citenamefont {Montgomery}, \citenamefont {Rezaei},
  \citenamefont {Ruisard}, \citenamefont {Sutter}, \citenamefont {Zhang} \emph
  {et~al.}}]{Bernal:2016_recirculator}%
  \BibitemOpen
  \bibfield  {author} {\bibinfo {author} {\bibfnamefont {S.}~\bibnamefont
  {Bernal}}, \bibinfo {author} {\bibfnamefont {B.}~\bibnamefont {Beaudoin}},
  \bibinfo {author} {\bibfnamefont {I.}~\bibnamefont {Haber}}, \bibinfo
  {author} {\bibfnamefont {T.}~\bibnamefont {Koeth}}, \bibinfo {author}
  {\bibfnamefont {Y.}~\bibnamefont {Mo}}, \bibinfo {author} {\bibfnamefont
  {E.}~\bibnamefont {Montgomery}}, \bibinfo {author} {\bibfnamefont
  {K.}~\bibnamefont {Rezaei}}, \bibinfo {author} {\bibfnamefont
  {K.}~\bibnamefont {Ruisard}}, \bibinfo {author} {\bibfnamefont
  {D.}~\bibnamefont {Sutter}}, \bibinfo {author} {\bibfnamefont
  {H.}~\bibnamefont {Zhang}},  \emph {et~al.},\ }in\ \href@noop {} {\emph
  {\bibinfo {booktitle} {AIP Conference Proceedings}}},\ Vol.\ \bibinfo
  {volume} {1777}\ (\bibinfo {organization} {AIP Publishing},\ \bibinfo {year}
  {2016})\ p.\ \bibinfo {pages} {100003}\BibitemShut {NoStop}%
\bibitem [{\citenamefont {Bulanov}\ \emph {et~al.}(2002)\citenamefont
  {Bulanov}, \citenamefont {Esirkepov}, \citenamefont {Kamenets}, \citenamefont
  {Kato}, \citenamefont {Kuznetsov}, \citenamefont {Nishihara}, \citenamefont
  {Pegoraro}, \citenamefont {Tajima},\ and\ \citenamefont
  {Khoroshkov}}]{Bulanov:2002_charged_beam_generation}%
  \BibitemOpen
  \bibfield  {author} {\bibinfo {author} {\bibfnamefont {S.}~\bibnamefont
  {Bulanov}}, \bibinfo {author} {\bibfnamefont {T.~Z.}\ \bibnamefont
  {Esirkepov}}, \bibinfo {author} {\bibfnamefont {F.}~\bibnamefont {Kamenets}},
  \bibinfo {author} {\bibfnamefont {Y.}~\bibnamefont {Kato}}, \bibinfo {author}
  {\bibfnamefont {A.}~\bibnamefont {Kuznetsov}}, \bibinfo {author}
  {\bibfnamefont {K.}~\bibnamefont {Nishihara}}, \bibinfo {author}
  {\bibfnamefont {F.}~\bibnamefont {Pegoraro}}, \bibinfo {author}
  {\bibfnamefont {T.}~\bibnamefont {Tajima}}, \ and\ \bibinfo {author}
  {\bibfnamefont {V.}~\bibnamefont {Khoroshkov}},\ }\href@noop {} {\bibfield
  {journal} {\bibinfo  {journal} {Plasma Physics Reports}\ }\textbf {\bibinfo
  {volume} {28}},\ \bibinfo {pages} {975} (\bibinfo {year} {2002})}\BibitemShut
  {NoStop}%
\bibitem [{\citenamefont {Fukuda}\ \emph {et~al.}(2009)\citenamefont {Fukuda},
  \citenamefont {Faenov}, \citenamefont {Tampo}, \citenamefont {Pikuz},
  \citenamefont {Nakamura}, \citenamefont {Kando}, \citenamefont {Hayashi},
  \citenamefont {Yogo}, \citenamefont {Sakaki}, \citenamefont {Kameshima} \emph
  {et~al.}}]{Fukuda:2009_species_generation}%
  \BibitemOpen
  \bibfield  {author} {\bibinfo {author} {\bibfnamefont {Y.}~\bibnamefont
  {Fukuda}}, \bibinfo {author} {\bibfnamefont {A.~Y.}\ \bibnamefont {Faenov}},
  \bibinfo {author} {\bibfnamefont {M.}~\bibnamefont {Tampo}}, \bibinfo
  {author} {\bibfnamefont {T.}~\bibnamefont {Pikuz}}, \bibinfo {author}
  {\bibfnamefont {T.}~\bibnamefont {Nakamura}}, \bibinfo {author}
  {\bibfnamefont {M.}~\bibnamefont {Kando}}, \bibinfo {author} {\bibfnamefont
  {Y.}~\bibnamefont {Hayashi}}, \bibinfo {author} {\bibfnamefont
  {A.}~\bibnamefont {Yogo}}, \bibinfo {author} {\bibfnamefont {H.}~\bibnamefont
  {Sakaki}}, \bibinfo {author} {\bibfnamefont {T.}~\bibnamefont {Kameshima}},
  \emph {et~al.},\ }\href@noop {} {\bibfield  {journal} {\bibinfo  {journal}
  {Physical review letters}\ }\textbf {\bibinfo {volume} {103}},\ \bibinfo
  {pages} {165002} (\bibinfo {year} {2009})}\BibitemShut {NoStop}%
\bibitem [{\citenamefont {Esirkepov}\ \emph {et~al.}(2004)\citenamefont
  {Esirkepov}, \citenamefont {Borghesi}, \citenamefont {Bulanov}, \citenamefont
  {Mourou},\ and\ \citenamefont
  {Tajima}}]{Esirkepov:2004_highly_efficient_ion_generation}%
  \BibitemOpen
  \bibfield  {author} {\bibinfo {author} {\bibfnamefont {T.}~\bibnamefont
  {Esirkepov}}, \bibinfo {author} {\bibfnamefont {M.}~\bibnamefont {Borghesi}},
  \bibinfo {author} {\bibfnamefont {S.}~\bibnamefont {Bulanov}}, \bibinfo
  {author} {\bibfnamefont {G.}~\bibnamefont {Mourou}}, \ and\ \bibinfo {author}
  {\bibfnamefont {T.}~\bibnamefont {Tajima}},\ }\href@noop {} {\bibfield
  {journal} {\bibinfo  {journal} {Physical review letters}\ }\textbf {\bibinfo
  {volume} {92}},\ \bibinfo {pages} {175003} (\bibinfo {year}
  {2004})}\BibitemShut {NoStop}%
\bibitem [{\citenamefont {Kaplan}(2015)}]{Kaplan:2015_preprint}%
  \BibitemOpen
  \bibfield  {author} {\bibinfo {author} {\bibfnamefont {A.}~\bibnamefont
  {Kaplan}},\ }\href@noop {} {\bibfield  {journal} {\bibinfo  {journal} {arXiv
  preprint arXiv:1503.06368}\ } (\bibinfo {year} {2015})}\BibitemShut {NoStop}%
\bibitem [{\citenamefont {Parks}\ \emph {et~al.}(2001)\citenamefont {Parks},
  \citenamefont {Cowan}, \citenamefont {Stephens},\ and\ \citenamefont
  {Campbell}}]{Parks:2001_neutron_production}%
  \BibitemOpen
  \bibfield  {author} {\bibinfo {author} {\bibfnamefont {P.}~\bibnamefont
  {Parks}}, \bibinfo {author} {\bibfnamefont {T.}~\bibnamefont {Cowan}},
  \bibinfo {author} {\bibfnamefont {R.}~\bibnamefont {Stephens}}, \ and\
  \bibinfo {author} {\bibfnamefont {E.}~\bibnamefont {Campbell}},\ }\href@noop
  {} {\bibfield  {journal} {\bibinfo  {journal} {Physical Review A}\ }\textbf
  {\bibinfo {volume} {63}},\ \bibinfo {pages} {063203} (\bibinfo {year}
  {2001})}\BibitemShut {NoStop}%
\bibitem [{\citenamefont {Bychenkov}\ \emph {et~al.}(2015)\citenamefont
  {Bychenkov}, \citenamefont {Brantov}, \citenamefont {Govras},\ and\
  \citenamefont {Kovalev}}]{Bychenkov_2015_review}%
  \BibitemOpen
  \bibfield  {author} {\bibinfo {author} {\bibfnamefont {V.~Y.}\ \bibnamefont
  {Bychenkov}}, \bibinfo {author} {\bibfnamefont {A.~V.}\ \bibnamefont
  {Brantov}}, \bibinfo {author} {\bibfnamefont {E.~A.}\ \bibnamefont {Govras}},
  \ and\ \bibinfo {author} {\bibfnamefont {V.~F.}\ \bibnamefont {Kovalev}},\
  }\href@noop {} {\bibfield  {journal} {\bibinfo  {journal} {Physics-Uspekhi}\
  }\textbf {\bibinfo {volume} {58}},\ \bibinfo {pages} {71} (\bibinfo {year}
  {2015})}\BibitemShut {NoStop}%
\bibitem [{\citenamefont {Murphy}\ \emph {et~al.}(2014)\citenamefont {Murphy},
  \citenamefont {Speirs}, \citenamefont {Sheludko}, \citenamefont {Putkunz},
  \citenamefont {McCulloch}, \citenamefont {Sparkes},\ and\ \citenamefont
  {Scholten}}]{Murphy:2014_cold_ions}%
  \BibitemOpen
  \bibfield  {author} {\bibinfo {author} {\bibfnamefont {D.}~\bibnamefont
  {Murphy}}, \bibinfo {author} {\bibfnamefont {R.}~\bibnamefont {Speirs}},
  \bibinfo {author} {\bibfnamefont {D.}~\bibnamefont {Sheludko}}, \bibinfo
  {author} {\bibfnamefont {C.}~\bibnamefont {Putkunz}}, \bibinfo {author}
  {\bibfnamefont {A.}~\bibnamefont {McCulloch}}, \bibinfo {author}
  {\bibfnamefont {B.}~\bibnamefont {Sparkes}}, \ and\ \bibinfo {author}
  {\bibfnamefont {R.}~\bibnamefont {Scholten}},\ }\href@noop {} {\bibfield
  {journal} {\bibinfo  {journal} {Nature communications}\ }\textbf {\bibinfo
  {volume} {5}} (\bibinfo {year} {2014})}\BibitemShut {NoStop}%
\bibitem [{\citenamefont {Gahlmann}\ \emph {et~al.}(2008)\citenamefont
  {Gahlmann}, \citenamefont {Park},\ and\ \citenamefont
  {Zewail}}]{Gahlmann:2008_ultrashort}%
  \BibitemOpen
  \bibfield  {author} {\bibinfo {author} {\bibfnamefont {A.}~\bibnamefont
  {Gahlmann}}, \bibinfo {author} {\bibfnamefont {S.~T.}\ \bibnamefont {Park}},
  \ and\ \bibinfo {author} {\bibfnamefont {A.~H.}\ \bibnamefont {Zewail}},\
  }\href@noop {} {\bibfield  {journal} {\bibinfo  {journal} {Physical Chemistry
  Chemical Physics}\ }\textbf {\bibinfo {volume} {10}},\ \bibinfo {pages}
  {2894} (\bibinfo {year} {2008})}\BibitemShut {NoStop}%
\bibitem [{\citenamefont {Booske}\ \emph {et~al.}(2011)\citenamefont {Booske},
  \citenamefont {Dobbs}, \citenamefont {Joye}, \citenamefont {Kory},
  \citenamefont {Neil}, \citenamefont {Park}, \citenamefont {Park},\ and\
  \citenamefont {Temkin}}]{Booske:2011_vacuum_review}%
  \BibitemOpen
  \bibfield  {author} {\bibinfo {author} {\bibfnamefont {J.~H.}\ \bibnamefont
  {Booske}}, \bibinfo {author} {\bibfnamefont {R.~J.}\ \bibnamefont {Dobbs}},
  \bibinfo {author} {\bibfnamefont {C.~D.}\ \bibnamefont {Joye}}, \bibinfo
  {author} {\bibfnamefont {C.~L.}\ \bibnamefont {Kory}}, \bibinfo {author}
  {\bibfnamefont {G.~R.}\ \bibnamefont {Neil}}, \bibinfo {author}
  {\bibfnamefont {G.-S.}\ \bibnamefont {Park}}, \bibinfo {author}
  {\bibfnamefont {J.}~\bibnamefont {Park}}, \ and\ \bibinfo {author}
  {\bibfnamefont {R.~J.}\ \bibnamefont {Temkin}},\ }\href@noop {} {\bibfield
  {journal} {\bibinfo  {journal} {IEEE Transactions on Terahertz Science and
  Technology}\ }\textbf {\bibinfo {volume} {1}},\ \bibinfo {pages} {54}
  (\bibinfo {year} {2011})}\BibitemShut {NoStop}%
\bibitem [{\citenamefont {Liu}\ \emph {et~al.}(2015)\citenamefont {Liu},
  \citenamefont {Zhang}, \citenamefont {Chen},\ and\ \citenamefont
  {Ang}}]{Liu:2015_maximal_charge}%
  \BibitemOpen
  \bibfield  {author} {\bibinfo {author} {\bibfnamefont {Y.}~\bibnamefont
  {Liu}}, \bibinfo {author} {\bibfnamefont {P.}~\bibnamefont {Zhang}}, \bibinfo
  {author} {\bibfnamefont {S.}~\bibnamefont {Chen}}, \ and\ \bibinfo {author}
  {\bibfnamefont {L.}~\bibnamefont {Ang}},\ }\href@noop {} {\bibfield
  {journal} {\bibinfo  {journal} {Physical Review Special Topics-Accelerators
  and Beams}\ }\textbf {\bibinfo {volume} {18}},\ \bibinfo {pages} {123402}
  (\bibinfo {year} {2015})}\BibitemShut {NoStop}%
\bibitem [{\citenamefont {Zhang}\ and\ \citenamefont
  {Lau}(2016)}]{Zhang:2016_review}%
  \BibitemOpen
  \bibfield  {author} {\bibinfo {author} {\bibfnamefont {P.}~\bibnamefont
  {Zhang}}\ and\ \bibinfo {author} {\bibfnamefont {Y.}~\bibnamefont {Lau}},\
  }\href@noop {} {\bibfield  {journal} {\bibinfo  {journal} {Journal of Plasma
  Physics}\ }\textbf {\bibinfo {volume} {82}} (\bibinfo {year}
  {2016})}\BibitemShut {NoStop}%
\bibitem [{\citenamefont {Jansen}(1988)}]{Jansen:1988_book}%
  \BibitemOpen
  \bibfield  {author} {\bibinfo {author} {\bibfnamefont {G.~H.}\ \bibnamefont
  {Jansen}},\ }\href@noop {} {\emph {\bibinfo {title} {Coulomb interactions in
  particle beams [book]}}}\ (\bibinfo  {publisher} {Technische Universiteit
  Delft},\ \bibinfo {address} {Delft},\ \bibinfo {year} {1988})\BibitemShut
  {NoStop}%
\bibitem [{\citenamefont {Reiser}(1994)}]{Reiser:1994_book}%
  \BibitemOpen
  \bibfield  {author} {\bibinfo {author} {\bibfnamefont {M.}~\bibnamefont
  {Reiser}},\ }\href@noop {} {\emph {\bibinfo {title} {Theory and Design of
  Charged Particle Beams}}}\ (\bibinfo  {publisher} {John Wiley \& Sons},\
  \bibinfo {address} {New York},\ \bibinfo {year} {1994})\BibitemShut {NoStop}%
\bibitem [{\citenamefont {Batygin}(2001)}]{Batygin:2001_self}%
  \BibitemOpen
  \bibfield  {author} {\bibinfo {author} {\bibfnamefont {Y.~K.}\ \bibnamefont
  {Batygin}},\ }\href@noop {} {\bibfield  {journal} {\bibinfo  {journal}
  {Physics of Plasmas}\ }\textbf {\bibinfo {volume} {8}},\ \bibinfo {pages}
  {3103} (\bibinfo {year} {2001})}\BibitemShut {NoStop}%
\bibitem [{\citenamefont {Bychenkov}\ and\ \citenamefont
  {Kovalev}(2005)}]{Bychenkov:2005_coulomb_explosion}%
  \BibitemOpen
  \bibfield  {author} {\bibinfo {author} {\bibfnamefont {V.~Y.}\ \bibnamefont
  {Bychenkov}}\ and\ \bibinfo {author} {\bibfnamefont {V.}~\bibnamefont
  {Kovalev}},\ }\href@noop {} {\bibfield  {journal} {\bibinfo  {journal}
  {Plasma physics reports}\ }\textbf {\bibinfo {volume} {31}},\ \bibinfo
  {pages} {178} (\bibinfo {year} {2005})}\BibitemShut {NoStop}%
\bibitem [{\citenamefont {Grech}\ \emph {et~al.}(2011)\citenamefont {Grech},
  \citenamefont {Nuter}, \citenamefont {Mikaberidze}, \citenamefont
  {Di~Cintio}, \citenamefont {Gremillet}, \citenamefont {Lefebvre},
  \citenamefont {Saalmann}, \citenamefont {Rost},\ and\ \citenamefont
  {Skupin}}]{Grech:2011_coulomb_explosion}%
  \BibitemOpen
  \bibfield  {author} {\bibinfo {author} {\bibfnamefont {M.}~\bibnamefont
  {Grech}}, \bibinfo {author} {\bibfnamefont {R.}~\bibnamefont {Nuter}},
  \bibinfo {author} {\bibfnamefont {A.}~\bibnamefont {Mikaberidze}}, \bibinfo
  {author} {\bibfnamefont {P.}~\bibnamefont {Di~Cintio}}, \bibinfo {author}
  {\bibfnamefont {L.}~\bibnamefont {Gremillet}}, \bibinfo {author}
  {\bibfnamefont {E.}~\bibnamefont {Lefebvre}}, \bibinfo {author}
  {\bibfnamefont {U.}~\bibnamefont {Saalmann}}, \bibinfo {author}
  {\bibfnamefont {J.~M.}\ \bibnamefont {Rost}}, \ and\ \bibinfo {author}
  {\bibfnamefont {S.}~\bibnamefont {Skupin}},\ }\href@noop {} {\bibfield
  {journal} {\bibinfo  {journal} {Physical Review E}\ }\textbf {\bibinfo
  {volume} {84}},\ \bibinfo {pages} {056404} (\bibinfo {year}
  {2011})}\BibitemShut {NoStop}%
\bibitem [{\citenamefont {Kaplan}\ \emph {et~al.}(2003)\citenamefont {Kaplan},
  \citenamefont {Dubetsky},\ and\ \citenamefont
  {Shkolnikov}}]{Kaplan:2003_shock}%
  \BibitemOpen
  \bibfield  {author} {\bibinfo {author} {\bibfnamefont {A.~E.}\ \bibnamefont
  {Kaplan}}, \bibinfo {author} {\bibfnamefont {B.~Y.}\ \bibnamefont
  {Dubetsky}}, \ and\ \bibinfo {author} {\bibfnamefont {P.}~\bibnamefont
  {Shkolnikov}},\ }\href@noop {} {\bibfield  {journal} {\bibinfo  {journal}
  {Physical review letters}\ }\textbf {\bibinfo {volume} {91}},\ \bibinfo
  {pages} {143401} (\bibinfo {year} {2003})}\BibitemShut {NoStop}%
\bibitem [{\citenamefont {Kovalev}\ and\ \citenamefont
  {Bychenkov}(2005)}]{Kovalev:2005_kinetic_spherically_coulomb_explosion}%
  \BibitemOpen
  \bibfield  {author} {\bibinfo {author} {\bibfnamefont {V.}~\bibnamefont
  {Kovalev}}\ and\ \bibinfo {author} {\bibfnamefont {V.~Y.}\ \bibnamefont
  {Bychenkov}},\ }\href@noop {} {\bibfield  {journal} {\bibinfo  {journal}
  {Journal of Experimental and Theoretical Physics}\ }\textbf {\bibinfo
  {volume} {101}},\ \bibinfo {pages} {212} (\bibinfo {year}
  {2005})}\BibitemShut {NoStop}%
\bibitem [{\citenamefont {Last}\ \emph {et~al.}(1997)\citenamefont {Last},
  \citenamefont {Schek},\ and\ \citenamefont
  {Jortner}}]{Last:1997_analytic_coulomb_explosion}%
  \BibitemOpen
  \bibfield  {author} {\bibinfo {author} {\bibfnamefont {I.}~\bibnamefont
  {Last}}, \bibinfo {author} {\bibfnamefont {I.}~\bibnamefont {Schek}}, \ and\
  \bibinfo {author} {\bibfnamefont {J.}~\bibnamefont {Jortner}},\ }\href@noop
  {} {\bibfield  {journal} {\bibinfo  {journal} {The Journal of chemical
  physics}\ }\textbf {\bibinfo {volume} {107}},\ \bibinfo {pages} {6685}
  (\bibinfo {year} {1997})}\BibitemShut {NoStop}%
\bibitem [{\citenamefont {Eloy}\ \emph {et~al.}(2001)\citenamefont {Eloy},
  \citenamefont {Azambuja}, \citenamefont {Mendonca},\ and\ \citenamefont
  {Bingham}}]{Eloy:2001_coulomb_explosion}%
  \BibitemOpen
  \bibfield  {author} {\bibinfo {author} {\bibfnamefont {M.}~\bibnamefont
  {Eloy}}, \bibinfo {author} {\bibfnamefont {R.}~\bibnamefont {Azambuja}},
  \bibinfo {author} {\bibfnamefont {J.}~\bibnamefont {Mendonca}}, \ and\
  \bibinfo {author} {\bibfnamefont {R.}~\bibnamefont {Bingham}},\ }\href@noop
  {} {\bibfield  {journal} {\bibinfo  {journal} {Physics of Plasmas}\ }\textbf
  {\bibinfo {volume} {8}},\ \bibinfo {pages} {1084} (\bibinfo {year}
  {2001})}\BibitemShut {NoStop}%
\bibitem [{\citenamefont {Krainov}\ and\ \citenamefont
  {Roshchupkin}(2001)}]{Krainov:2001_ce_dynamics}%
  \BibitemOpen
  \bibfield  {author} {\bibinfo {author} {\bibfnamefont {V.}~\bibnamefont
  {Krainov}}\ and\ \bibinfo {author} {\bibfnamefont {A.}~\bibnamefont
  {Roshchupkin}},\ }\href@noop {} {\bibfield  {journal} {\bibinfo  {journal}
  {Physical Review A}\ }\textbf {\bibinfo {volume} {64}},\ \bibinfo {pages}
  {063204} (\bibinfo {year} {2001})}\BibitemShut {NoStop}%
\bibitem [{\citenamefont {Morrison}\ and\ \citenamefont
  {Grant}(2015)}]{Morrison:2015_slow_down_dynamics}%
  \BibitemOpen
  \bibfield  {author} {\bibinfo {author} {\bibfnamefont {J.}~\bibnamefont
  {Morrison}}\ and\ \bibinfo {author} {\bibfnamefont {E.}~\bibnamefont
  {Grant}},\ }\href@noop {} {\bibfield  {journal} {\bibinfo  {journal}
  {Physical Review A}\ }\textbf {\bibinfo {volume} {91}},\ \bibinfo {pages}
  {023423} (\bibinfo {year} {2015})}\BibitemShut {NoStop}%
\bibitem [{\citenamefont {Boella}\ \emph {et~al.}(2016)\citenamefont {Boella},
  \citenamefont {Paradisi}, \citenamefont {D’Angola}, \citenamefont {Silva},\
  and\ \citenamefont {Coppa}}]{Boella:2016_multiple_species}%
  \BibitemOpen
  \bibfield  {author} {\bibinfo {author} {\bibfnamefont {E.}~\bibnamefont
  {Boella}}, \bibinfo {author} {\bibfnamefont {B.~P.}\ \bibnamefont
  {Paradisi}}, \bibinfo {author} {\bibfnamefont {A.}~\bibnamefont
  {D’Angola}}, \bibinfo {author} {\bibfnamefont {L.~O.}\ \bibnamefont
  {Silva}}, \ and\ \bibinfo {author} {\bibfnamefont {G.}~\bibnamefont
  {Coppa}},\ }\href@noop {} {\bibfield  {journal} {\bibinfo  {journal} {Journal
  of Plasma Physics}\ }\textbf {\bibinfo {volume} {82}},\ \bibinfo {pages}
  {905820110} (\bibinfo {year} {2016})}\BibitemShut {NoStop}%
\bibitem [{\citenamefont {Degtyareva}\ \emph {et~al.}(1998)\citenamefont
  {Degtyareva}, \citenamefont {Monastyrsky}, \citenamefont {Schelev},\ and\
  \citenamefont {Tarasov}}]{Degtyareva:1998_gaussian_pileup}%
  \BibitemOpen
  \bibfield  {author} {\bibinfo {author} {\bibfnamefont {V.~P.}\ \bibnamefont
  {Degtyareva}}, \bibinfo {author} {\bibfnamefont {M.~A.}\ \bibnamefont
  {Monastyrsky}}, \bibinfo {author} {\bibfnamefont {M.~Y.}\ \bibnamefont
  {Schelev}}, \ and\ \bibinfo {author} {\bibfnamefont {V.~A.}\ \bibnamefont
  {Tarasov}},\ }\href@noop {} {\bibfield  {journal} {\bibinfo  {journal}
  {Optical Engineering}\ }\textbf {\bibinfo {volume} {37}},\ \bibinfo {pages}
  {2227} (\bibinfo {year} {1998})}\BibitemShut {NoStop}%
\bibitem [{\citenamefont {Siwick}\ \emph {et~al.}(2002)\citenamefont {Siwick},
  \citenamefont {Dwyer}, \citenamefont {Jordan},\ and\ \citenamefont
  {Dwayne~Miller}}]{Siwick:2002_mean_field}%
  \BibitemOpen
  \bibfield  {author} {\bibinfo {author} {\bibfnamefont {B.~J.}\ \bibnamefont
  {Siwick}}, \bibinfo {author} {\bibfnamefont {J.~R.}\ \bibnamefont {Dwyer}},
  \bibinfo {author} {\bibfnamefont {R.~E.}\ \bibnamefont {Jordan}}, \ and\
  \bibinfo {author} {\bibfnamefont {R.~J.}\ \bibnamefont {Dwayne~Miller}},\
  }\href@noop {} {\bibfield  {journal} {\bibinfo  {journal} {Journal of Applied
  Physics}\ }\textbf {\bibinfo {volume} {92}},\ \bibinfo {pages} {1643}
  (\bibinfo {year} {2002})}\BibitemShut {NoStop}%
\bibitem [{\citenamefont {Qian}\ and\ \citenamefont
  {Elsayed-Ali}(2002)}]{Qian:2002_fluid_flow}%
  \BibitemOpen
  \bibfield  {author} {\bibinfo {author} {\bibfnamefont {B.-L.}\ \bibnamefont
  {Qian}}\ and\ \bibinfo {author} {\bibfnamefont {H.~E.}\ \bibnamefont
  {Elsayed-Ali}},\ }\href@noop {} {\bibfield  {journal} {\bibinfo  {journal}
  {Journal of Applied Physics}\ }\textbf {\bibinfo {volume} {91}},\ \bibinfo
  {pages} {462} (\bibinfo {year} {2002})}\BibitemShut {NoStop}%
\bibitem [{\citenamefont {Reed}(2006)}]{Reed:2006_short_pulse_theory}%
  \BibitemOpen
  \bibfield  {author} {\bibinfo {author} {\bibfnamefont {B.~W.}\ \bibnamefont
  {Reed}},\ }\href@noop {} {\bibfield  {journal} {\bibinfo  {journal} {Journal
  of Applied Physics}\ }\textbf {\bibinfo {volume} {100}},\ \bibinfo {pages}
  {034916} (\bibinfo {year} {2006})}\BibitemShut {NoStop}%
\bibitem [{\citenamefont {Collin}\ \emph {et~al.}(2005)\citenamefont {Collin},
  \citenamefont {Merano}, \citenamefont {Gatri}, \citenamefont {Sonderegger},
  \citenamefont {Renucci}, \citenamefont {Ganiere},\ and\ \citenamefont
  {Deveaud}}]{Collin:2005_broadening}%
  \BibitemOpen
  \bibfield  {author} {\bibinfo {author} {\bibfnamefont {S.}~\bibnamefont
  {Collin}}, \bibinfo {author} {\bibfnamefont {M.}~\bibnamefont {Merano}},
  \bibinfo {author} {\bibfnamefont {M.}~\bibnamefont {Gatri}}, \bibinfo
  {author} {\bibfnamefont {S.}~\bibnamefont {Sonderegger}}, \bibinfo {author}
  {\bibfnamefont {P.}~\bibnamefont {Renucci}}, \bibinfo {author} {\bibfnamefont
  {J.-D.}\ \bibnamefont {Ganiere}}, \ and\ \bibinfo {author} {\bibfnamefont
  {B.}~\bibnamefont {Deveaud}},\ }\href@noop {} {\bibfield  {journal} {\bibinfo
   {journal} {Journal of applied physics}\ }\textbf {\bibinfo {volume} {98}},\
  \bibinfo {pages} {094910} (\bibinfo {year} {2005})}\BibitemShut {NoStop}%
\bibitem [{\citenamefont {Tao}\ \emph {et~al.}(2012)\citenamefont {Tao},
  \citenamefont {Zhang}, \citenamefont {Duxbury}, \citenamefont {Berz},\ and\
  \citenamefont {Ruan}}]{Tao:2012_space_charge}%
  \BibitemOpen
  \bibfield  {author} {\bibinfo {author} {\bibfnamefont {Z.}~\bibnamefont
  {Tao}}, \bibinfo {author} {\bibfnamefont {H.}~\bibnamefont {Zhang}}, \bibinfo
  {author} {\bibfnamefont {P.}~\bibnamefont {Duxbury}}, \bibinfo {author}
  {\bibfnamefont {M.}~\bibnamefont {Berz}}, \ and\ \bibinfo {author}
  {\bibfnamefont {C.-Y.}\ \bibnamefont {Ruan}},\ }\href@noop {} {\bibfield
  {journal} {\bibinfo  {journal} {Journal of Applied Physics}\ }\textbf
  {\bibinfo {volume} {111}},\ \bibinfo {pages} {044316} (\bibinfo {year}
  {2012})}\BibitemShut {NoStop}%
\bibitem [{\citenamefont {Portman}\ \emph {et~al.}(2013)\citenamefont
  {Portman}, \citenamefont {Zhang}, \citenamefont {Tao}, \citenamefont
  {Makino}, \citenamefont {Berz}, \citenamefont {Duxbury},\ and\ \citenamefont
  {Ruan}}]{Portman:2013_computational_characterization}%
  \BibitemOpen
  \bibfield  {author} {\bibinfo {author} {\bibfnamefont {J.}~\bibnamefont
  {Portman}}, \bibinfo {author} {\bibfnamefont {H.}~\bibnamefont {Zhang}},
  \bibinfo {author} {\bibfnamefont {Z.}~\bibnamefont {Tao}}, \bibinfo {author}
  {\bibfnamefont {K.}~\bibnamefont {Makino}}, \bibinfo {author} {\bibfnamefont
  {M.}~\bibnamefont {Berz}}, \bibinfo {author} {\bibfnamefont {P.}~\bibnamefont
  {Duxbury}}, \ and\ \bibinfo {author} {\bibfnamefont {C.-Y.}\ \bibnamefont
  {Ruan}},\ }\href@noop {} {\bibfield  {journal} {\bibinfo  {journal} {Applied
  Physics Letters}\ }\textbf {\bibinfo {volume} {103}},\ \bibinfo {pages}
  {253115} (\bibinfo {year} {2013})}\BibitemShut {NoStop}%
\bibitem [{\citenamefont {Portman}\ \emph {et~al.}(2014)\citenamefont
  {Portman}, \citenamefont {Zhang}, \citenamefont {Makino}, \citenamefont
  {Ruan}, \citenamefont {Berz},\ and\ \citenamefont
  {Duxbury}}]{Portman:2014_image_charge}%
  \BibitemOpen
  \bibfield  {author} {\bibinfo {author} {\bibfnamefont {J.}~\bibnamefont
  {Portman}}, \bibinfo {author} {\bibfnamefont {H.}~\bibnamefont {Zhang}},
  \bibinfo {author} {\bibfnamefont {K.}~\bibnamefont {Makino}}, \bibinfo
  {author} {\bibfnamefont {C.}~\bibnamefont {Ruan}}, \bibinfo {author}
  {\bibfnamefont {M.}~\bibnamefont {Berz}}, \ and\ \bibinfo {author}
  {\bibfnamefont {P.}~\bibnamefont {Duxbury}},\ }\href@noop {} {\bibfield
  {journal} {\bibinfo  {journal} {Journal of Applied Physics}\ }\textbf
  {\bibinfo {volume} {116}},\ \bibinfo {pages} {174302} (\bibinfo {year}
  {2014})}\BibitemShut {NoStop}%
\bibitem [{\citenamefont {Michalik}\ and\ \citenamefont
  {Sipe}(2006)}]{Michalik:2006_analytic_gaussian}%
  \BibitemOpen
  \bibfield  {author} {\bibinfo {author} {\bibfnamefont {A.}~\bibnamefont
  {Michalik}}\ and\ \bibinfo {author} {\bibfnamefont {J.}~\bibnamefont
  {Sipe}},\ }\href@noop {} {\bibfield  {journal} {\bibinfo  {journal} {Journal
  of applied physics}\ }\textbf {\bibinfo {volume} {99}},\ \bibinfo {pages}
  {054908} (\bibinfo {year} {2006})}\BibitemShut {NoStop}%
\bibitem [{\citenamefont {King}\ \emph {et~al.}(2005)\citenamefont {King},
  \citenamefont {Campbell}, \citenamefont {Frank}, \citenamefont {Reed},
  \citenamefont {Schmerge}, \citenamefont {Siwick}, \citenamefont {Stuart},\
  and\ \citenamefont {Weber}}]{King:2005_review}%
  \BibitemOpen
  \bibfield  {author} {\bibinfo {author} {\bibfnamefont {W.~E.}\ \bibnamefont
  {King}}, \bibinfo {author} {\bibfnamefont {G.~H.}\ \bibnamefont {Campbell}},
  \bibinfo {author} {\bibfnamefont {A.}~\bibnamefont {Frank}}, \bibinfo
  {author} {\bibfnamefont {B.}~\bibnamefont {Reed}}, \bibinfo {author}
  {\bibfnamefont {J.~F.}\ \bibnamefont {Schmerge}}, \bibinfo {author}
  {\bibfnamefont {B.~J.}\ \bibnamefont {Siwick}}, \bibinfo {author}
  {\bibfnamefont {B.~C.}\ \bibnamefont {Stuart}}, \ and\ \bibinfo {author}
  {\bibfnamefont {P.~M.}\ \bibnamefont {Weber}},\ }\href@noop {} {\bibfield
  {journal} {\bibinfo  {journal} {Journal of Applied Physics}\ }\textbf
  {\bibinfo {volume} {97}},\ \bibinfo {pages} {111101} (\bibinfo {year}
  {2005})}\BibitemShut {NoStop}%
\bibitem [{\citenamefont {Hall}\ \emph {et~al.}(2014)\citenamefont {Hall},
  \citenamefont {Stemmer}, \citenamefont {Zheng}, \citenamefont {Zhu},\ and\
  \citenamefont {Maracas}}]{Hall:2014_report}%
  \BibitemOpen
  \bibfield  {author} {\bibinfo {author} {\bibfnamefont {E.}~\bibnamefont
  {Hall}}, \bibinfo {author} {\bibfnamefont {S.}~\bibnamefont {Stemmer}},
  \bibinfo {author} {\bibfnamefont {H.}~\bibnamefont {Zheng}}, \bibinfo
  {author} {\bibfnamefont {Y.}~\bibnamefont {Zhu}}, \ and\ \bibinfo {author}
  {\bibfnamefont {G.}~\bibnamefont {Maracas}},\ }\href@noop {} {\emph {\bibinfo
  {title} {Future of Electron Scattering and Diffraction}}},\ \bibinfo {type}
  {Tech. Rep.}\ (\bibinfo  {institution} {US Department of Energy, Washington,
  DC (United States)},\ \bibinfo {year} {2014})\BibitemShut {NoStop}%
\bibitem [{\citenamefont {Williams}\ \emph
  {et~al.}(2017{\natexlab{a}})\citenamefont {Williams}, \citenamefont {Zhou},
  \citenamefont {Sun}, \citenamefont {Chang}, \citenamefont {Makino},
  \citenamefont {Berz}, \citenamefont {Duxbury},\ and\ \citenamefont
  {Ruan}}]{Williams:2017_longitudinal_emittance}%
  \BibitemOpen
  \bibfield  {author} {\bibinfo {author} {\bibfnamefont {J.}~\bibnamefont
  {Williams}}, \bibinfo {author} {\bibfnamefont {F.}~\bibnamefont {Zhou}},
  \bibinfo {author} {\bibfnamefont {T.}~\bibnamefont {Sun}}, \bibinfo {author}
  {\bibfnamefont {K.}~\bibnamefont {Chang}}, \bibinfo {author} {\bibfnamefont
  {K.}~\bibnamefont {Makino}}, \bibinfo {author} {\bibfnamefont
  {M.}~\bibnamefont {Berz}}, \bibinfo {author} {\bibfnamefont {P.}~\bibnamefont
  {Duxbury}}, \ and\ \bibinfo {author} {\bibfnamefont {C.-Y.}\ \bibnamefont
  {Ruan}},\ }\href@noop {} {\bibfield  {journal} {\bibinfo  {journal}
  {Structural Dynamics}\ }\textbf {\bibinfo {volume} {4}} (\bibinfo {year}
  {2017}{\natexlab{a}})}\BibitemShut {NoStop}%
\bibitem [{\citenamefont {Valfells}\ \emph {et~al.}(2002)\citenamefont
  {Valfells}, \citenamefont {Feldman}, \citenamefont {Virgo}, \citenamefont
  {O'shea},\ and\ \citenamefont {Lau}}]{Valfells:2002_vc_limit}%
  \BibitemOpen
  \bibfield  {author} {\bibinfo {author} {\bibfnamefont {A.}~\bibnamefont
  {Valfells}}, \bibinfo {author} {\bibfnamefont {D.}~\bibnamefont {Feldman}},
  \bibinfo {author} {\bibfnamefont {M.}~\bibnamefont {Virgo}}, \bibinfo
  {author} {\bibfnamefont {P.}~\bibnamefont {O'shea}}, \ and\ \bibinfo {author}
  {\bibfnamefont {Y.}~\bibnamefont {Lau}},\ }\href@noop {} {\bibfield
  {journal} {\bibinfo  {journal} {Physics of Plasmas (1994-present)}\ }\textbf
  {\bibinfo {volume} {9}},\ \bibinfo {pages} {2377} (\bibinfo {year}
  {2002})}\BibitemShut {NoStop}%
\bibitem [{\citenamefont {Luiten}\ \emph {et~al.}(2004)\citenamefont {Luiten},
  \citenamefont {Geer}, \citenamefont {Loos}, \citenamefont {Kiewiet},\ and\
  \citenamefont {Wiel}}]{Luiten:2004_uniform_ellipsoidal}%
  \BibitemOpen
  \bibfield  {author} {\bibinfo {author} {\bibfnamefont {O.~J.}\ \bibnamefont
  {Luiten}}, \bibinfo {author} {\bibfnamefont {S.~B. V.~D.}\ \bibnamefont
  {Geer}}, \bibinfo {author} {\bibfnamefont {M.~J.~D.}\ \bibnamefont {Loos}},
  \bibinfo {author} {\bibfnamefont {F.~B.}\ \bibnamefont {Kiewiet}}, \ and\
  \bibinfo {author} {\bibfnamefont {M.~J. V.~D.}\ \bibnamefont {Wiel}},\
  }\href@noop {} {\bibfield  {journal} {\bibinfo  {journal} {Physical review
  letters}\ }\textbf {\bibinfo {volume} {93}},\ \bibinfo {pages} {094802}
  (\bibinfo {year} {2004})}\BibitemShut {NoStop}%
\bibitem [{\citenamefont {Miller}(2014)}]{Miller:2014_science_review}%
  \BibitemOpen
  \bibfield  {author} {\bibinfo {author} {\bibfnamefont {R.~D.}\ \bibnamefont
  {Miller}},\ }\href@noop {} {\bibfield  {journal} {\bibinfo  {journal}
  {Science}\ }\textbf {\bibinfo {volume} {343}},\ \bibinfo {pages} {1108}
  (\bibinfo {year} {2014})}\BibitemShut {NoStop}%
\bibitem [{\citenamefont {Srinivasan}\ \emph {et~al.}(2003)\citenamefont
  {Srinivasan}, \citenamefont {Lobastov}, \citenamefont {Ruan},\ and\
  \citenamefont {Zewail}}]{Srinivasan:2003_UED}%
  \BibitemOpen
  \bibfield  {author} {\bibinfo {author} {\bibfnamefont {R.}~\bibnamefont
  {Srinivasan}}, \bibinfo {author} {\bibfnamefont {V.}~\bibnamefont
  {Lobastov}}, \bibinfo {author} {\bibfnamefont {C.-Y.}\ \bibnamefont {Ruan}},
  \ and\ \bibinfo {author} {\bibfnamefont {A.}~\bibnamefont {Zewail}},\
  }\href@noop {} {\bibfield  {journal} {\bibinfo  {journal} {Helvetica Chimica
  Acta}\ }\textbf {\bibinfo {volume} {86}},\ \bibinfo {pages} {1761} (\bibinfo
  {year} {2003})}\BibitemShut {NoStop}%
\bibitem [{\citenamefont {Ruan}\ \emph {et~al.}(2009)\citenamefont {Ruan},
  \citenamefont {Murooka}, \citenamefont {Raman}, \citenamefont {Murdick},
  \citenamefont {Worhatch},\ and\ \citenamefont
  {Pell}}]{Ruan:2009_nanocrystallography}%
  \BibitemOpen
  \bibfield  {author} {\bibinfo {author} {\bibfnamefont {C.-Y.}\ \bibnamefont
  {Ruan}}, \bibinfo {author} {\bibfnamefont {Y.}~\bibnamefont {Murooka}},
  \bibinfo {author} {\bibfnamefont {R.~K.}\ \bibnamefont {Raman}}, \bibinfo
  {author} {\bibfnamefont {R.~A.}\ \bibnamefont {Murdick}}, \bibinfo {author}
  {\bibfnamefont {R.~J.}\ \bibnamefont {Worhatch}}, \ and\ \bibinfo {author}
  {\bibfnamefont {A.}~\bibnamefont {Pell}},\ }\href@noop {} {\bibfield
  {journal} {\bibinfo  {journal} {Microscopy and Microanalysis}\ }\textbf
  {\bibinfo {volume} {15}},\ \bibinfo {pages} {323} (\bibinfo {year}
  {2009})}\BibitemShut {NoStop}%
\bibitem [{\citenamefont {Van~Oudheusden}\ \emph {et~al.}(2010)\citenamefont
  {Van~Oudheusden}, \citenamefont {Pasmans}, \citenamefont {Van Der~Geer},
  \citenamefont {De~Loos}, \citenamefont {Van Der~Wiel},\ and\ \citenamefont
  {Luiten}}]{van_Oudheusden:2010_rf_compression_experiment}%
  \BibitemOpen
  \bibfield  {author} {\bibinfo {author} {\bibfnamefont {T.}~\bibnamefont
  {Van~Oudheusden}}, \bibinfo {author} {\bibfnamefont {P.}~\bibnamefont
  {Pasmans}}, \bibinfo {author} {\bibfnamefont {S.}~\bibnamefont {Van
  Der~Geer}}, \bibinfo {author} {\bibfnamefont {M.}~\bibnamefont {De~Loos}},
  \bibinfo {author} {\bibfnamefont {M.}~\bibnamefont {Van Der~Wiel}}, \ and\
  \bibinfo {author} {\bibfnamefont {O.}~\bibnamefont {Luiten}},\ }\href@noop {}
  {\bibfield  {journal} {\bibinfo  {journal} {Physical review letters}\
  }\textbf {\bibinfo {volume} {105}},\ \bibinfo {pages} {264801} (\bibinfo
  {year} {2010})}\BibitemShut {NoStop}%
\bibitem [{\citenamefont {Sciaini}\ and\ \citenamefont
  {Miller}(2011)}]{Sciaini:2011_review}%
  \BibitemOpen
  \bibfield  {author} {\bibinfo {author} {\bibfnamefont {G.}~\bibnamefont
  {Sciaini}}\ and\ \bibinfo {author} {\bibfnamefont {R.~D.}\ \bibnamefont
  {Miller}},\ }\href@noop {} {\bibfield  {journal} {\bibinfo  {journal}
  {Reports on Progress in Physics}\ }\textbf {\bibinfo {volume} {74}},\
  \bibinfo {pages} {096101} (\bibinfo {year} {2011})}\BibitemShut {NoStop}%
\bibitem [{\citenamefont {Musumeci}\ \emph {et~al.}(2010)\citenamefont
  {Musumeci}, \citenamefont {Moody}, \citenamefont {Scoby}, \citenamefont
  {Gutierrez}, \citenamefont {Westfall},\ and\ \citenamefont
  {Li}}]{Musumeci:2010_single_shot}%
  \BibitemOpen
  \bibfield  {author} {\bibinfo {author} {\bibfnamefont {P.}~\bibnamefont
  {Musumeci}}, \bibinfo {author} {\bibfnamefont {J.}~\bibnamefont {Moody}},
  \bibinfo {author} {\bibfnamefont {C.}~\bibnamefont {Scoby}}, \bibinfo
  {author} {\bibfnamefont {M.}~\bibnamefont {Gutierrez}}, \bibinfo {author}
  {\bibfnamefont {M.}~\bibnamefont {Westfall}}, \ and\ \bibinfo {author}
  {\bibfnamefont {R.}~\bibnamefont {Li}},\ }\href@noop {} {\bibfield  {journal}
  {\bibinfo  {journal} {Journal of Applied Physics}\ }\textbf {\bibinfo
  {volume} {108}},\ \bibinfo {pages} {114513} (\bibinfo {year}
  {2010})}\BibitemShut {NoStop}%
\bibitem [{\citenamefont {Weathersby}\ \emph {et~al.}(2015)\citenamefont
  {Weathersby}, \citenamefont {Brown}, \citenamefont {Centurion}, \citenamefont
  {Chase}, \citenamefont {Coffee}, \citenamefont {Corbett}, \citenamefont
  {Eichner}, \citenamefont {Frisch}, \citenamefont {Fry}, \citenamefont
  {G{\"u}hr}, \citenamefont {Hartmann}, \citenamefont {Hast}, \citenamefont
  {Hettel}, \citenamefont {Jobe}, \citenamefont {Jongewaard}, \citenamefont
  {Lewandowski}, \citenamefont {Li}, \citenamefont {Lindenberg}, \citenamefont
  {Makasyuk}, \citenamefont {May}, \citenamefont {McCormick}, \citenamefont
  {Nguyen}, \citenamefont {Reid}, \citenamefont {Shen}, \citenamefont
  {Sokolowski-Tinten}, \citenamefont {Vecchione}, \citenamefont {Vetter},
  \citenamefont {Wu}, \citenamefont {Yang}, \citenamefont {A.},\ and\
  \citenamefont {Wang}}]{Weathersby:2015_slac}%
  \BibitemOpen
  \bibfield  {author} {\bibinfo {author} {\bibfnamefont {S.}~\bibnamefont
  {Weathersby}}, \bibinfo {author} {\bibfnamefont {G.}~\bibnamefont {Brown}},
  \bibinfo {author} {\bibfnamefont {M.}~\bibnamefont {Centurion}}, \bibinfo
  {author} {\bibfnamefont {T.}~\bibnamefont {Chase}}, \bibinfo {author}
  {\bibfnamefont {R.}~\bibnamefont {Coffee}}, \bibinfo {author} {\bibfnamefont
  {J.}~\bibnamefont {Corbett}}, \bibinfo {author} {\bibfnamefont
  {J.}~\bibnamefont {Eichner}}, \bibinfo {author} {\bibfnamefont
  {J.}~\bibnamefont {Frisch}}, \bibinfo {author} {\bibfnamefont
  {A.}~\bibnamefont {Fry}}, \bibinfo {author} {\bibfnamefont {M.}~\bibnamefont
  {G{\"u}hr}}, \bibinfo {author} {\bibfnamefont {N.}~\bibnamefont {Hartmann}},
  \bibinfo {author} {\bibfnamefont {C.}~\bibnamefont {Hast}}, \bibinfo {author}
  {\bibfnamefont {R.}~\bibnamefont {Hettel}}, \bibinfo {author} {\bibfnamefont
  {R.~K.}\ \bibnamefont {Jobe}}, \bibinfo {author} {\bibfnamefont {E.~N.}\
  \bibnamefont {Jongewaard}}, \bibinfo {author} {\bibfnamefont {J.~R.}\
  \bibnamefont {Lewandowski}}, \bibinfo {author} {\bibfnamefont {R.~K.}\
  \bibnamefont {Li}}, \bibinfo {author} {\bibfnamefont {A.~M.}\ \bibnamefont
  {Lindenberg}}, \bibinfo {author} {\bibfnamefont {I.}~\bibnamefont
  {Makasyuk}}, \bibinfo {author} {\bibfnamefont {J.~E.}\ \bibnamefont {May}},
  \bibinfo {author} {\bibfnamefont {D.}~\bibnamefont {McCormick}}, \bibinfo
  {author} {\bibfnamefont {M.~N.}\ \bibnamefont {Nguyen}}, \bibinfo {author}
  {\bibfnamefont {A.~H.}\ \bibnamefont {Reid}}, \bibinfo {author}
  {\bibfnamefont {X.}~\bibnamefont {Shen}}, \bibinfo {author} {\bibfnamefont
  {K.}~\bibnamefont {Sokolowski-Tinten}}, \bibinfo {author} {\bibfnamefont
  {T.}~\bibnamefont {Vecchione}}, \bibinfo {author} {\bibfnamefont {S.~L.}\
  \bibnamefont {Vetter}}, \bibinfo {author} {\bibfnamefont {J.}~\bibnamefont
  {Wu}}, \bibinfo {author} {\bibfnamefont {J.}~\bibnamefont {Yang}}, \bibinfo
  {author} {\bibfnamefont {D.~H.}\ \bibnamefont {A.}}, \ and\ \bibinfo {author}
  {\bibfnamefont {X.~J.}\ \bibnamefont {Wang}},\ }\href@noop {} {\bibfield
  {journal} {\bibinfo  {journal} {Review of Scientific Instruments}\ }\textbf
  {\bibinfo {volume} {86}},\ \bibinfo {pages} {073702} (\bibinfo {year}
  {2015})}\BibitemShut {NoStop}%
\bibitem [{\citenamefont {Murooka}\ \emph {et~al.}(2011)\citenamefont
  {Murooka}, \citenamefont {Naruse}, \citenamefont {Sakakihara}, \citenamefont
  {Ishimaru}, \citenamefont {Yang},\ and\ \citenamefont
  {Tanimura}}]{Murooka:2011_TED}%
  \BibitemOpen
  \bibfield  {author} {\bibinfo {author} {\bibfnamefont {Y.}~\bibnamefont
  {Murooka}}, \bibinfo {author} {\bibfnamefont {N.}~\bibnamefont {Naruse}},
  \bibinfo {author} {\bibfnamefont {S.}~\bibnamefont {Sakakihara}}, \bibinfo
  {author} {\bibfnamefont {M.}~\bibnamefont {Ishimaru}}, \bibinfo {author}
  {\bibfnamefont {J.}~\bibnamefont {Yang}}, \ and\ \bibinfo {author}
  {\bibfnamefont {K.}~\bibnamefont {Tanimura}},\ }\href@noop {} {\bibfield
  {journal} {\bibinfo  {journal} {Applied Physics Letters}\ }\textbf {\bibinfo
  {volume} {98}},\ \bibinfo {pages} {251903} (\bibinfo {year}
  {2011})}\BibitemShut {NoStop}%
\bibitem [{\citenamefont {Rosenzweig}\ \emph {et~al.}(2006)\citenamefont
  {Rosenzweig}, \citenamefont {Cook}, \citenamefont {England}, \citenamefont
  {Dunning}, \citenamefont {Anderson},\ and\ \citenamefont
  {Ferrario}}]{Rosenzweig:2006_emittance_compensation}%
  \BibitemOpen
  \bibfield  {author} {\bibinfo {author} {\bibfnamefont {J.}~\bibnamefont
  {Rosenzweig}}, \bibinfo {author} {\bibfnamefont {A.}~\bibnamefont {Cook}},
  \bibinfo {author} {\bibfnamefont {R.}~\bibnamefont {England}}, \bibinfo
  {author} {\bibfnamefont {M.}~\bibnamefont {Dunning}}, \bibinfo {author}
  {\bibfnamefont {S.}~\bibnamefont {Anderson}}, \ and\ \bibinfo {author}
  {\bibfnamefont {M.}~\bibnamefont {Ferrario}},\ }\href@noop {} {\bibfield
  {journal} {\bibinfo  {journal} {Nuclear Instruments and Methods in Physics
  Research Section A: Accelerators, Spectrometers, Detectors and Associated
  Equipment}\ }\textbf {\bibinfo {volume} {557}},\ \bibinfo {pages} {87}
  (\bibinfo {year} {2006})}\BibitemShut {NoStop}%
\bibitem [{\citenamefont {Musumeci}\ \emph {et~al.}(2008)\citenamefont
  {Musumeci}, \citenamefont {Moody}, \citenamefont {England}, \citenamefont
  {Rosenzweig},\ and\ \citenamefont
  {Tran}}]{Musucemi:2008_generate_uniform_ellipsoid}%
  \BibitemOpen
  \bibfield  {author} {\bibinfo {author} {\bibfnamefont {P.}~\bibnamefont
  {Musumeci}}, \bibinfo {author} {\bibfnamefont {J.~T.}\ \bibnamefont {Moody}},
  \bibinfo {author} {\bibfnamefont {R.~J.}\ \bibnamefont {England}}, \bibinfo
  {author} {\bibfnamefont {J.~B.}\ \bibnamefont {Rosenzweig}}, \ and\ \bibinfo
  {author} {\bibfnamefont {T.}~\bibnamefont {Tran}},\ }\href@noop {} {\bibfield
   {journal} {\bibinfo  {journal} {Physical review letters}\ }\textbf {\bibinfo
  {volume} {100}},\ \bibinfo {pages} {244801} (\bibinfo {year}
  {2008})}\BibitemShut {NoStop}%
\bibitem [{\citenamefont {Williams}\ \emph
  {et~al.}(2017{\natexlab{b}})\citenamefont {Williams}, \citenamefont {Zhou},
  \citenamefont {Sun}, \citenamefont {Duxbury}, \citenamefont {Lund},
  \citenamefont {Zerbe},\ and\ \citenamefont
  {Ruan}}]{Williams:2017_transverse_emittance}%
  \BibitemOpen
  \bibfield  {author} {\bibinfo {author} {\bibfnamefont {J.}~\bibnamefont
  {Williams}}, \bibinfo {author} {\bibfnamefont {F.}~\bibnamefont {Zhou}},
  \bibinfo {author} {\bibfnamefont {T.}~\bibnamefont {Sun}}, \bibinfo {author}
  {\bibfnamefont {P.}~\bibnamefont {Duxbury}}, \bibinfo {author} {\bibfnamefont
  {S.}~\bibnamefont {Lund}}, \bibinfo {author} {\bibfnamefont {B.}~\bibnamefont
  {Zerbe}}, \ and\ \bibinfo {author} {\bibfnamefont {C.-Y.}\ \bibnamefont
  {Ruan}},\ }\href@noop {} {\bibfield  {journal} {\bibinfo  {journal} {Bulletin
  of the American Physical Society}\ }\textbf {\bibinfo {volume} {62}}
  (\bibinfo {year} {2017}{\natexlab{b}})}\BibitemShut {NoStop}%
\bibitem [{\citenamefont {Berz}\ \emph {et~al.}(1987)\citenamefont {Berz},
  \citenamefont {Hoffmann},\ and\ \citenamefont {Wollnik}}]{Berz:1987_cosy}%
  \BibitemOpen
  \bibfield  {author} {\bibinfo {author} {\bibfnamefont {M.}~\bibnamefont
  {Berz}}, \bibinfo {author} {\bibfnamefont {H.}~\bibnamefont {Hoffmann}}, \
  and\ \bibinfo {author} {\bibfnamefont {H.}~\bibnamefont {Wollnik}},\
  }\href@noop {} {\bibfield  {journal} {\bibinfo  {journal} {Nuclear
  Instruments and Methods in Physics Research Section A: Accelerators,
  Spectrometers, Detectors and Associated Equipment}\ }\textbf {\bibinfo
  {volume} {258}},\ \bibinfo {pages} {402} (\bibinfo {year}
  {1987})}\BibitemShut {NoStop}%
\bibitem [{\citenamefont {Zhang}\ \emph {et~al.}(2015)\citenamefont {Zhang},
  \citenamefont {Portman}, \citenamefont {Tao}, \citenamefont {Duxbury},
  \citenamefont {Ruan}, \citenamefont {Makino},\ and\ \citenamefont
  {Berz}}]{Zhang:2015_fmm_cosy}%
  \BibitemOpen
  \bibfield  {author} {\bibinfo {author} {\bibfnamefont {H.}~\bibnamefont
  {Zhang}}, \bibinfo {author} {\bibfnamefont {J.}~\bibnamefont {Portman}},
  \bibinfo {author} {\bibfnamefont {Z.}~\bibnamefont {Tao}}, \bibinfo {author}
  {\bibfnamefont {P.}~\bibnamefont {Duxbury}}, \bibinfo {author} {\bibfnamefont
  {C.-Y.}\ \bibnamefont {Ruan}}, \bibinfo {author} {\bibfnamefont
  {K.}~\bibnamefont {Makino}}, \ and\ \bibinfo {author} {\bibfnamefont
  {M.}~\bibnamefont {Berz}},\ }\href@noop {} {\bibfield  {journal} {\bibinfo
  {journal} {Microscopy and Microanalysis}\ }\textbf {\bibinfo {volume} {21}},\
  \bibinfo {pages} {224} (\bibinfo {year} {2015})}\BibitemShut {NoStop}%
\bibitem [{\citenamefont {Friedman}\ \emph {et~al.}(2014)\citenamefont
  {Friedman}, \citenamefont {Cohen}, \citenamefont {Grote}, \citenamefont
  {Lund}, \citenamefont {Sharp}, \citenamefont {Vay}, \citenamefont {Haber},\
  and\ \citenamefont {Kishek}}]{Friedman:2014_warp}%
  \BibitemOpen
  \bibfield  {author} {\bibinfo {author} {\bibfnamefont {A.}~\bibnamefont
  {Friedman}}, \bibinfo {author} {\bibfnamefont {R.~H.}\ \bibnamefont {Cohen}},
  \bibinfo {author} {\bibfnamefont {D.~P.}\ \bibnamefont {Grote}}, \bibinfo
  {author} {\bibfnamefont {S.~M.}\ \bibnamefont {Lund}}, \bibinfo {author}
  {\bibfnamefont {W.~M.}\ \bibnamefont {Sharp}}, \bibinfo {author}
  {\bibfnamefont {J.-L.}\ \bibnamefont {Vay}}, \bibinfo {author} {\bibfnamefont
  {I.}~\bibnamefont {Haber}}, \ and\ \bibinfo {author} {\bibfnamefont {R.~A.}\
  \bibnamefont {Kishek}},\ }\href@noop {} {\bibfield  {journal} {\bibinfo
  {journal} {IEEE Transactions on Plasma Science}\ }\textbf {\bibinfo {volume}
  {42}},\ \bibinfo {pages} {1321} (\bibinfo {year} {2014})}\BibitemShut
  {NoStop}%
\bibitem [{\citenamefont {Morrison}\ \emph {et~al.}(2013)\citenamefont
  {Morrison}, \citenamefont {Chatelain}, \citenamefont {Godbout},\ and\
  \citenamefont {Siwick}}]{Morrison:2013_measurement}%
  \BibitemOpen
  \bibfield  {author} {\bibinfo {author} {\bibfnamefont {V.~R.}\ \bibnamefont
  {Morrison}}, \bibinfo {author} {\bibfnamefont {R.~P.}\ \bibnamefont
  {Chatelain}}, \bibinfo {author} {\bibfnamefont {C.}~\bibnamefont {Godbout}},
  \ and\ \bibinfo {author} {\bibfnamefont {B.~J.}\ \bibnamefont {Siwick}},\
  }\href@noop {} {\bibfield  {journal} {\bibinfo  {journal} {Optics express}\
  }\textbf {\bibinfo {volume} {21}},\ \bibinfo {pages} {21} (\bibinfo {year}
  {2013})}\BibitemShut {NoStop}%
\bibitem [{\citenamefont {Li}\ and\ \citenamefont
  {Lewellen}(2008)}]{Li:2008_quasiellipsoidal}%
  \BibitemOpen
  \bibfield  {author} {\bibinfo {author} {\bibfnamefont {Y.}~\bibnamefont
  {Li}}\ and\ \bibinfo {author} {\bibfnamefont {J.~W.}\ \bibnamefont
  {Lewellen}},\ }\href@noop {} {\bibfield  {journal} {\bibinfo  {journal}
  {Physical review letters}\ }\textbf {\bibinfo {volume} {100}},\ \bibinfo
  {pages} {074801} (\bibinfo {year} {2008})}\BibitemShut {NoStop}%
\bibitem [{\citenamefont {Lessner}\ \emph {et~al.}(2016)\citenamefont
  {Lessner}, \citenamefont {Wang},\ and\ \citenamefont
  {Musumeci}}]{BES_report:2016_electron_sources}%
  \BibitemOpen
  \bibfield  {author} {\bibinfo {author} {\bibfnamefont {E.}~\bibnamefont
  {Lessner}}, \bibinfo {author} {\bibfnamefont {X.}~\bibnamefont {Wang}}, \
  and\ \bibinfo {author} {\bibfnamefont {P.}~\bibnamefont {Musumeci}},\
  }\href@noop {} {\emph {\bibinfo {title} {Report of the Basic Energy Sciences
  Workshop on the Future of Electron Sources}}},\ \bibinfo {type} {~}\
  (\bibinfo  {institution} {SLAC National Accelerator Laboratory},\ \bibinfo
  {year} {2016})\BibitemShut {NoStop}%
\bibitem [{\citenamefont {Boyer}\ \emph {et~al.}(1989)\citenamefont {Boyer},
  \citenamefont {Luk}, \citenamefont {Solem},\ and\ \citenamefont
  {Rhodes}}]{Boyer:1989_kinetic_energy}%
  \BibitemOpen
  \bibfield  {author} {\bibinfo {author} {\bibfnamefont {K.}~\bibnamefont
  {Boyer}}, \bibinfo {author} {\bibfnamefont {T.}~\bibnamefont {Luk}}, \bibinfo
  {author} {\bibfnamefont {J.}~\bibnamefont {Solem}}, \ and\ \bibinfo {author}
  {\bibfnamefont {C.}~\bibnamefont {Rhodes}},\ }\href@noop {} {\bibfield
  {journal} {\bibinfo  {journal} {Physical Review A}\ }\textbf {\bibinfo
  {volume} {39}},\ \bibinfo {pages} {1186} (\bibinfo {year}
  {1989})}\BibitemShut {NoStop}%
\bibitem [{\citenamefont {Wangler}\ \emph {et~al.}(1985)\citenamefont
  {Wangler}, \citenamefont {Crandall}, \citenamefont {Mills},\ and\
  \citenamefont {Reiser}}]{Wangler:1985_emittance_relaxation}%
  \BibitemOpen
  \bibfield  {author} {\bibinfo {author} {\bibfnamefont {T.}~\bibnamefont
  {Wangler}}, \bibinfo {author} {\bibfnamefont {K.}~\bibnamefont {Crandall}},
  \bibinfo {author} {\bibfnamefont {R.}~\bibnamefont {Mills}}, \ and\ \bibinfo
  {author} {\bibfnamefont {M.}~\bibnamefont {Reiser}},\ }\href@noop {}
  {\bibfield  {journal} {\bibinfo  {journal} {IEEE Transactions on Nuclear
  Science}\ }\textbf {\bibinfo {volume} {32}},\ \bibinfo {pages} {2196}
  (\bibinfo {year} {1985})}\BibitemShut {NoStop}%
\bibitem [{\citenamefont {Luginsland}\ \emph {et~al.}(1996)\citenamefont
  {Luginsland}, \citenamefont {Lau},\ and\ \citenamefont
  {Gilgenbach}}]{Luginsland:1996_child_langmuir_2d}%
  \BibitemOpen
  \bibfield  {author} {\bibinfo {author} {\bibfnamefont {J.}~\bibnamefont
  {Luginsland}}, \bibinfo {author} {\bibfnamefont {Y.}~\bibnamefont {Lau}}, \
  and\ \bibinfo {author} {\bibfnamefont {R.}~\bibnamefont {Gilgenbach}},\
  }\href@noop {} {\bibfield  {journal} {\bibinfo  {journal} {Physical review
  letters}\ }\textbf {\bibinfo {volume} {77}},\ \bibinfo {pages} {4668}
  (\bibinfo {year} {1996})}\BibitemShut {NoStop}%
\bibitem [{\citenamefont {Gluckstern}(1994)}]{Gluckstern:1994_analytic_halo}%
  \BibitemOpen
  \bibfield  {author} {\bibinfo {author} {\bibfnamefont {R.~L.}\ \bibnamefont
  {Gluckstern}},\ }\href@noop {} {\bibfield  {journal} {\bibinfo  {journal}
  {Physical review letters}\ }\textbf {\bibinfo {volume} {73}},\ \bibinfo
  {pages} {1247} (\bibinfo {year} {1994})}\BibitemShut {NoStop}%
\bibitem [{\citenamefont {Wangler}\ \emph {et~al.}(1998)\citenamefont
  {Wangler}, \citenamefont {Crandall}, \citenamefont {Ryne},\ and\
  \citenamefont {Wang}}]{Wangler:1998_particle_core_review}%
  \BibitemOpen
  \bibfield  {author} {\bibinfo {author} {\bibfnamefont {T.}~\bibnamefont
  {Wangler}}, \bibinfo {author} {\bibfnamefont {K.}~\bibnamefont {Crandall}},
  \bibinfo {author} {\bibfnamefont {R.}~\bibnamefont {Ryne}}, \ and\ \bibinfo
  {author} {\bibfnamefont {T.}~\bibnamefont {Wang}},\ }\href@noop {} {\bibfield
   {journal} {\bibinfo  {journal} {Physical review special topics-accelerators
  and beams}\ }\textbf {\bibinfo {volume} {1}},\ \bibinfo {pages} {084201}
  (\bibinfo {year} {1998})}\BibitemShut {NoStop}%
\bibitem [{\citenamefont {Inc.}()}]{Mathematica}%
  \BibitemOpen
  \bibfield  {author} {\bibinfo {author} {\bibfnamefont {W.~R.}\ \bibnamefont
  {Inc.}},\ }\href@noop {} {\enquote {\bibinfo {title} {Mathematica, {V}ersion
  11.1},}\ }\bibinfo {note} {Champaign, IL, 2017}\BibitemShut {NoStop}%
\end{thebibliography}%

\end{document}